\documentclass[sigconf,nonacm]{acmart}
\usepackage{pvldb}

\renewcommand\vldbdoi{XX.XX/XXX.XX}
\renewcommand\vldbpages{XXX-XXX}
\renewcommand\vldbavailabilityurl{https://github.com/zpkhor/vldb27-flamingo-dag-bft}

\renewcommand\footnotetextcopyrightpermission[1]{}
\usepackage{savesym}
\usepackage[T1]{fontenc}
\restoresymbol{TXF}{iint}
\usepackage{amsfonts}
\usepackage{amsmath}
\usepackage{textcomp}
\usepackage{bbding}
\usepackage{xspace}
\usepackage{pifont}
\usepackage[normalem]{ulem}
\usepackage{hyphenat}
\usepackage{algorithm}
\usepackage{algorithmicx}
\usepackage[noend]{algpseudocode}
\usepackage{mathtools}
\usepackage{xspace}
\usepackage{enumitem}
\usepackage{multirow}
\usepackage{tikz}
\usepackage{pgfplots}
\usetikzlibrary{patterns}
\usepackage[dvipsnames]{xcolor} 
\usepackage{todonotes}          
\usepackage{algorithm}
\usepackage{algpseudocode}
\usepackage{amsmath}
\usepackage{graphicx}
\usepackage{subcaption}
\usepackage{multicol}
\usepackage[normalem]{ulem}

\newif\ifextend
\extendtrue

\algrenewcommand\algorithmicrequire{\textbf{Input:}}
\algrenewcommand\algorithmicensure{\textbf{Output:}}

\newcommand{\sys}{{Flamingo}\xspace}

\begin{document}
\title{\sys: On Load Balancing in DAG-based Consensus Protocols}

\author{Zhen Ping Khor}
\affiliation{
  \institution{University of Pennsylvania}
  \country{}
}
\email{zpkhor@upenn.edu}

\author{Garvit Gupta}
\affiliation{
  \institution{Apple}
  \country{}
}
\email{garvitgupta@icloud.com}

\author{Mohammad Javad Amiri}
\affiliation{
  \institution{Stony Brook University}
  \country{}
}
\email{amiri@cs.stonybrook.edu}

\author{Boon Thau Loo}
\affiliation{
  \institution{University of Pennsylvania}
  \country{}
}
\email{boonloo@seas.upenn.edu}

\fancyhead{}
\begin{abstract}
Distributed data management systems deployed in untrusted environments rely on Byzantine Fault-Tolerant (BFT) consensus protocols to tolerate malicious failures. DAG-based BFT protocols improve throughput by letting validators disseminate transactions concurrently and by scaling execution across multiple workers. However, imbalances in workload or resource capacity can still degrade performance significantly. This paper presents \sys, a load-balancing protocol for certified DAG-based BFT protocols that addresses imbalance at both the ordering and execution layers. 
At the ordering layer, \sys periodically migrates client accounts away from overloaded validators, adapting to skewed submissions and heterogeneous validator capacity while preserving correctness under Byzantine faults, with migrations taking effect only through the committed log.
At the execution layer, \sys redistributes committed transactions across executor workers using a deterministic, order-preserving scheduler that balances load and minimizes cross-worker data movement, without centralized coordination or costly distributed commit. Built on top of Narwhal and Tusk, our prototype shows that \sys recovers throughput and latency under workload skew, validator heterogeneity, and shifting hotspots, adds negligible overhead when the system is balanced, and needs load balancing in both layers, since resolving only one shifts the bottleneck to the other.
\end{abstract}

\maketitle
\section{Introduction}

Byzantine Fault-tolerant (BFT) protocols~\cite{yin2019hotstuff,gueta2019sbft,kotla2007zyzzyva,castro1999practical} allow distributed data management systems deployed in untrusted environments to tolerate malicious node behavior. In traditional BFT protocols, a designated leader receives client transactions and broadcasts them with an assigned order to all replicas. This ensures consistency but creates a resource imbalance, since the leader's storage, CPU, and bandwidth cap system throughput while the remaining replicas sit underutilized. Moreover, since dissemination and consensus are tightly coupled, the leader failure stalls the entire protocol.
\ifextend Rotating the leader~\cite{aiyer2005bar, kwon2014tendermint, yin2019hotstuff, chan2020streamlet, chan2018pala, chan2018pili, gilad2017algorand, hanke2018dfinity, kokoris2019robust, veronese2010ebawa, veronese2009spin, clement2009making, buchnik2020fireledger} distributes load more evenly across replicas over time~\cite{veronese2009spin, behl2015consensus, behl2017hybrids}, but introduces additional overhead due to more frequent view synchronization~\cite{yin2019hotstuff} and still requires every replica to be provisioned for peak leader load, since any replica may assume the leader role yet only one fully uses its capacity at a time.
\fi
Multi-proposer protocols~\cite{stathakopoulou2022mir, lyu2025ladon, arun2022scalable, lyu2025orthrus, gupta2021rcc} aim to mitigate this underutilization by running multiple BFT instances concurrently with each replica leading a parallel instance, but because dissemination and consensus remain tightly coupled, a single node's failure can still affect the entire protocol.

DAG-based consensus protocols~\cite{keidar2021all,dai2024remora,dai2023gradeddag,danezis2022narwhal,dai2024wahoo,spiegelman2024shoal,arun2025shoal++,cheng2024shardag,spiegelman2022bullshark,shrestha2024sailfish,malkhi2024bbca,jovanovic2024mahi,dai2024lightdag,raikwar2024sok,giridharan2024autobahn,nagda2026dag} address these limitations: validators disseminate transaction blocks concurrently, the blocks form a Directed Acyclic Graph; consensus is decoupled from dissemination, so node failures do not stall transaction processing. Each node consists of an ordering layer, a validator whose workers stream transaction batches in parallel, and an execution layer, whose executor workers each own a disjoint shard of the state; both layers scale horizontally by adding workers.

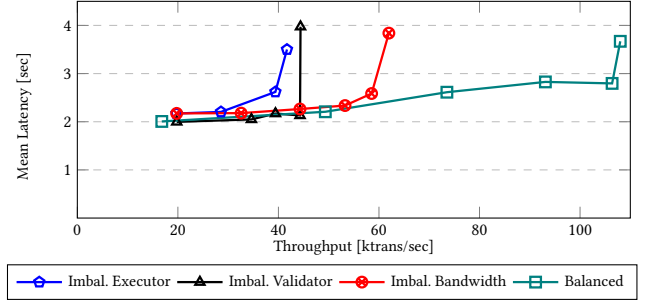
\begin{figure}[t]
\scriptsize
\centering
\begin{tikzpicture}[scale=1]
\begin{axis}[
    xlabel={Throughput [ktrans/sec]},
    xlabel style={yshift=1.5em},
    ylabel={Mean Latency [sec]},
    ylabel style={yshift=-2.5em,xshift=-0.4em},
    width=0.5\textwidth,
    height=.25\textwidth,
    xmin=0, xmax=110,
    ymin=0, ymax=4.5,
    xtick={0,20,40,60,80,100},
    ytick={1,2,3,4},
    legend columns=4, 
    legend style={
        at={(axis cs:-14,-1.7)},
        anchor=south west,
    },
    ymajorgrids=true,
    grid style=dashed,
]

\addplot[
    color=blue,
    mark=pentagon,
    mark size=2pt,
    line width=0.3mm,
    ]
    coordinates {
    (19.716,2.171)(28.574,2.204)(39.395,2.619)(41.7, 3.5)};

 \addplot[
    color=black,
    mark=triangle,
    mark size=2pt,
    line width=0.3mm,
    ]
    coordinates {
    (19.773,1.997)(34.622,2.048)(39.414,2.170)(44.297,2.131)(44.395,3.973)};

 \addplot[
    color=red,
    mark=otimes,
    mark size=2pt,
    line width=0.3mm,
    ]
    coordinates {
    (19.761,2.173)(32.585,2.179)(44.290,2.264)(53.251,2.336)(58.539,2.586)(61.971,3.840)};

 \addplot[
    color=teal,
    mark=square,
    mark size=2pt,
    line width=0.3mm,
    ]
    coordinates {
    (16.800,2.006)(49.330,2.207)(73.512,2.613)(93.103,2.827)(106.377,2.796)(108,3.67)};

\addlegendentry{Imbal. Executor}
\addlegendentry{Imbal. Validator}
\addlegendentry{Imbal. Bandwidth}
\addlegendentry{Balanced}
 
\end{axis}
\end{tikzpicture}
\caption{Impact of load imbalance and validator heterogeneity (a single node with limited bandwidth) on performance}
  \label{fig:motive}
\end{figure}

Despite their scalability, DAG-based protocols are highly sensitive to workload imbalance and validator heterogeneity. Figure~\ref{fig:motive} demonstrates this using the widely studied DAG-based BFT protocol Narwhal and Tusk~\cite{danezis2022narwhal} in a four-node local deployment. Under balanced resource capacity and load, the protocol achieves approximately $106$\,ktps. \ifextend
However, imbalances cause significant degradation: reducing one node's bandwidth to one-third of its peers lowers throughput to ${\sim}58$\,ktps ($45\%$ reduction), directing $90\%$ of transactions to a single validator (with the remaining $10\%$ split among three others) yields ${\sim}44$\,ktps ($58\%$ reduction), and concentrating $90\%$ of execution on a single executor worker drops throughput to ${\sim}39$\,ktps ($63\%$ reduction).
\fi These results motivate the need to address load and resource imbalances at both the ordering and execution layers to fully exploit the advantages of DAG-based protocols.

Balancing either layer is complicated by the setting itself. At the ordering layer, clients choose or are pre-assigned to validators, so skewed submissions saturate some validators while others idle, and heterogeneous hardware or fluctuating networks cause the same problem. Having no trusted coordinator makes the rebalancing more challenging: every validator must reach the same migration decision from its own, possibly divergent, DAG view, and a Byzantine validator may misreport its load to trigger harmful load balancing decisions. At the execution layer, skewed access patterns create hot shards, and distributed transactions spanning multiple executor workers require cross-worker coordination that degrades throughput even at low rates, as locks must be held during coordination~\cite{thomson2012calvin}. A replicated executor must additionally preserve the consensus-determined order deterministically, even across validators running different numbers of executor workers, which rules out schedulers that reorder batches for locality.

This paper presents \textit{\sys}, a load-balancing mechanism for DAG-based BFT protocols addressing imbalance at both ordering and execution layers while running off the critical path and introducing no new trust assumptions.

First, at the ordering layer, \sys balances load by changing which validator serves each client. Every client account starts with a validator derived deterministically, and \sys periodically migrates accounts from overloaded validators to those with spare capacity, prioritizing receivers with low latency to the client.
The load balancer runs on every validator and draws its load signals from certified metadata piggybacked on the DAG, requiring no dedicated metric exchange. Signals are distinguished by how they can be trusted: values that any peer can recompute from certified data determine whether and where load moves, but self-measured values can influence only how much load is migrated, never the safety of the underlying protocol.
A migration takes effect only through the committed log, requiring both committed evidence that $2f+1$ validators reached the same decision and the receiving validator's committed acceptance.
If the assigned validator becomes unresponsive, a bounded relay fallback lets a small deterministic set of validators admit the client's requests, keeping the account serviceable. We prove three properties of this migration layer: all honest validators agree on every account's assignment, every correct client's valid requests eventually commit, and fallback admission is bounded, so the fallback path cannot be abused to inject unbounded extra traffic.

Second, at the execution layer,  \sys schedules committed transactions across executor workers to balance load while minimizing the data movement required for distributed transactions. Every executor worker receives the full committed batch, which co-located LAN bandwidth readily absorbs, and independently runs the same lightweight deterministic scheduler, producing identical assignments without a central dispatcher or any synchronization of partition state. 
To perform transactions that access data hosted by multiple workers, \sys follows a data fusion model~\cite{lin2016towards}: the assigned worker pulls the remote inputs and performs all writes locally, a single one-way transfer instead of the gather-and-scatter round trip of existing protocols, e.g., PilotFish~\cite{kniep2025pilotfish}.
\ifextend It preserves the consensus order at lower scheduling algorithmic complexity compared to Hermes-style prescient routing~\cite{lin2021don} that reorders batches, which a replicated BFT executor cannot adopt.\fi

We have implemented \sys on top of Narwhal and Tusk~\cite{danezis2022narwhal}, the foundational certified-DAG design underlying industry-grade systems such as Sui~\cite{sui}. Since \sys does not alter transaction dissemination or the consensus routine, it can be incorporated with other certified DAG-based protocols such as Bullshark~\cite{spiegelman2022bullshark} and Shoal~\cite{spiegelman2024shoal}.
Our evaluation shows that \sys recovers throughput and latency under submission skew, heterogeneous validator bandwidth, and shifting hotspots. It also scales across validators and executor workers, reduces scheduling overhead compared with prescient routing~\cite{lin2021don}, and shows that both layers are needed to address their respective bottlenecks.
The prototype implements the load measurement and migration decision components of the ordering layer, while the Byzantine safety of the migration-commit lifecycle is established analytically.

Our contributions can be summarized as follows:

\begin{itemize}[leftmargin=1.2em,itemsep=1pt,topsep=2pt,parsep=0pt,partopsep=0pt]
    \item We identify and quantify the sensitivity of DAG-based BFT protocols to
    load imbalance, showing that submission skew, validator bandwidth
    heterogeneity, and executor hotspots reduce performance, and that the two affected layers, ordering and execution, both need to be addressed.
    \item At the ordering layer, we design an account-migration mechanism that derives its signals from certified DAG metadata and activates migrations only through committed evidence and receiver acceptance. We prove assignment agreement, liveness, and bounded fallback admission under Byzantine faults.
    \item At the execution layer, we design a deterministic, order-preserving, full-batch scheduler that combines load-aware placement with data fusion, reducing each distributed transaction to a single one-way state transfer.
    \item We implement a prototype of \sys and evaluate its performance across various imbalanced conditions, demonstrating throughput and latency recovery under imbalance with negligible overhead when the system is balanced.
\end{itemize}
\section{DAG-based BFT Consensus}\label{sec:background}\label{sec:dag-based}

Traditional BFT protocols funnel both transaction dissemination and ordering through a single leader, capping throughput at the leader's capacity while other replicas sit underutilized. A recent line of work decouples dissemination from the consensus routine and lets every node broadcast transaction blocks concurrently, organizing them into a Directed Acyclic Graph (DAG) whose edges reference preceding blocks~\cite{keidar2021all,dai2024remora,dai2023gradeddag,danezis2022narwhal,dai2024wahoo,spiegelman2024shoal,arun2025shoal++,cheng2024shardag,spiegelman2022bullshark,shrestha2024sailfish,malkhi2024bbca,jovanovic2024mahi,dai2024lightdag,raikwar2024sok,giridharan2024autobahn,nagda2026dag}. Processing a transaction then involves three steps: DAG construction, consensus, and execution. The DAG is built asynchronously to the total order, and each node maintains a local view; the consensus routine ensures a consistent view across nodes where each node derives the total order independently from its local DAG, without any additional communication phase. \ifextend Since consensus no longer blocks the flow of transactions and every replica builds and processes its DAG in parallel, these protocols utilize the resources of all nodes and achieve substantially higher throughput.\fi

\begin{figure}[t]
    \centering
    \includegraphics[width= 0.45\textwidth]{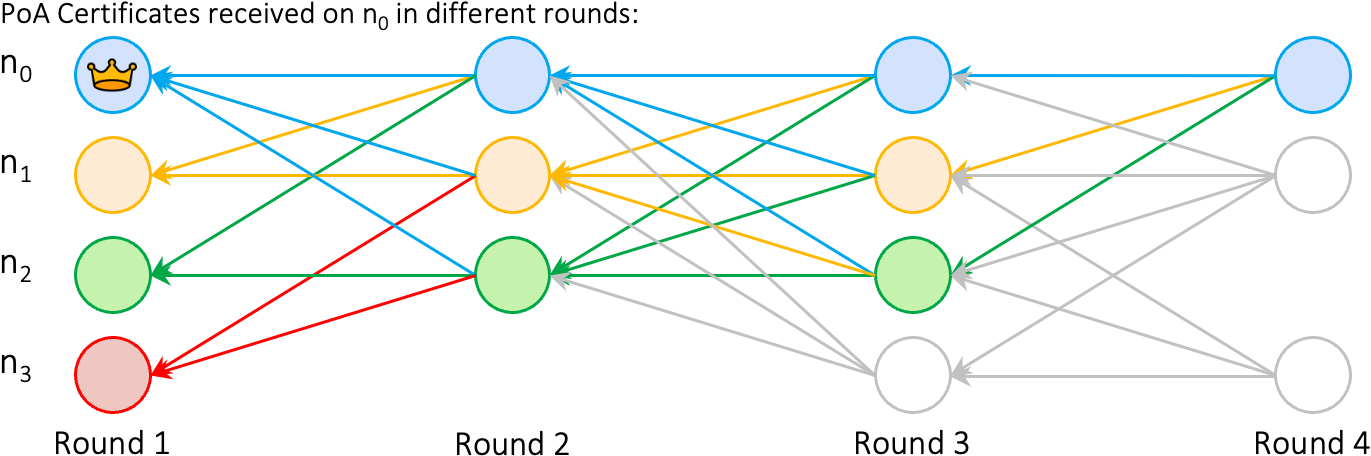}
    \caption{DAG construction on node $n_0$. Back edges reflect the PoA certificates that $n_0$ holds from the preceding round.
    \ifextend For example, $n_0$ receives certificates from $n_0$, $n_1$, and $n_2$ in round~1, so round~2 contains only these three vertices. When $n_0$ is elected as leader, all blocks reachable from the leader block in round~4 are committed.\fi}
    \label{fig:DAG-P0}
\end{figure}

Figure~\ref{fig:DAG-P0} illustrates DAG construction on node $n_0$ in a 4-node network using the Narwhal and Tusk protocols~\cite{danezis2022narwhal}. In each round, every node packs its client transactions into a block that carries the node's signature and its causal history, namely certificates of blocks from the preceding round, and broadcasts it. A receiving node validates and stores the block, then returns a signed acknowledgment over the block digest, round number, and creator identity. Once the creator collects $n-f$ distinct acknowledgments, it aggregates them into a Proof of Availability (PoA) certificate and broadcasts it, and all nodes include the certificates they receive in their subsequent blocks. A node advances to the next round after obtaining $n-f$ certificates for the current round, creating the corresponding vertices in its local DAG. For example, $n_0$ receives certificates from $n_0$, $n_1$, and $n_2$ in round 1 and thus creates only these three vertices in round 2; the back edges are determined by the certificates stored in the received blocks.
Since a PoA certificate carries $n-f$ signatures, at least $f+1$ honest validators have stored the block, so the block remains available for retrieval when needed for ordering, and quorum intersection prevents equivocation. \ifextend Because each block references certificates from the previous round, an inductive argument shows that the entire causal history is certified and available, satisfying causality.\fi

In the consensus phase, a node is (randomly) elected as leader for the final round of a wave (e.g., round 4), and each replica independently commits the blocks that have a path to the leader block in its local view, e.g., highlighted blocks in Figure~\ref{fig:DAG-P0}, yielding the same total order on all replicas without further communication. Blocks not yet ordered, e.g., gray blocks in Figure~\ref{fig:DAG-P0}, are committed in subsequent waves under newly elected leaders. Once the order is established, executor workers execute the transactions. DAG-based protocols typically shard the replicated state among executor workers, each executing the transactions that access its shard.

\ifextend Despite these gains, our experimental observations in Figure~\ref{fig:motive} reveal a critical shortcoming: performance is highly sensitive to balanced workloads and homogeneous deployments. When client load is skewed, or nodes differ in capacity, such as bandwidth, throughput and latency degrade sharply. This paper investigates and addresses this problem across the ordering and execution layers of DAG-based BFT protocols.\fi

\section{\sys Overview}
\label{sec:overview}

We assume an asynchronous network of $n = 3f + 1$ validators, of which at most $f$ may be Byzantine and exhibit arbitrary, potentially malicious behavior. The adversary is computationally bounded and cannot break cryptographic assumptions. Validators communicate over bidirectional point-to-point channels; a public-key infrastructure (PKI) with digital signatures ensures that messages from honest validators cannot be forged or repudiated, and a collision-resistant hash function $D(\cdot)$ produces fixed-size message digests.

Each node consists of a validator (ordering layer) and an executor (execution layer). The validator comprises a primary and several workers, which receive client transactions, seal and broadcast them as batches, and certify data availability through quorum acknowledgments. The execution layer partitions state across executor workers, which process committed batches in consensus-determined order.

\begin{figure}[t]
    \centering
    \includegraphics[width= 0.48\textwidth]{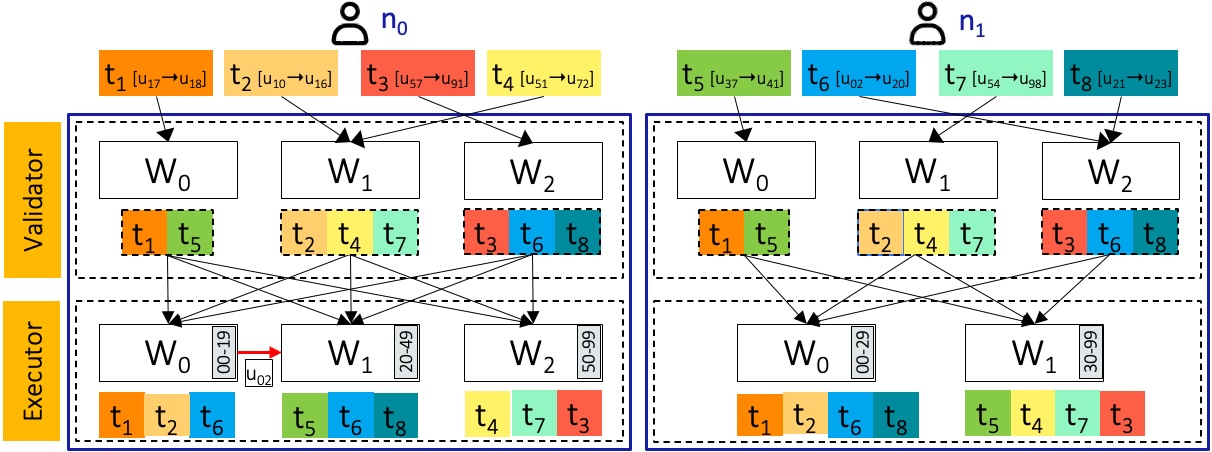}
    \caption{A partial view of \sys with nodes $n_0$ and $n_1$\ifextend, each consisting of 3 validator workers. Node $n_0$ has 3 executor workers, while $n_1$ includes 2 executor workers\fi.}
    \label{fig:arch}
\end{figure}

Figure~\ref{fig:arch} shows the architecture on $n_0$ and $n_1$. All nodes have the same number of validator workers, since each worker $w_i$ must have a counterpart $w_i$ at every other node to communicate with. The number of executor workers may differ (three for $n_0$, two for $n_1$), and each node may shard state differently. Each validator worker $w_i$ seals transactions from its assigned clients into batches, broadcasts them to its counterparts at the other nodes, and receives the batches those counterparts broadcast. For example, $n_0.w_0$ receives transaction $t_1$ while $n_1.w_0$ receives $t_5$, and each seals its transaction into a batch for the other.

Once a batch is committed, every node receives it, and each transaction is executed by the executor worker whose shard owns the accessed records. For instance, $n_0.w_0$ manages records $u_0$ to $u_{19}$ and executes transaction $t_1$, which accesses records $u_{17}$ and $u_{18}$. If a transaction, such as $t_6$, accesses records across different workers (e.g., $u_{02}$ on $n_0.w_0$ and $u_{20}$ on $n_0.w_1$), \sys ensures the relevant state is transferred among the workers.

\sys extends DAG-based protocols with two load-balancing mechanisms that run alongside consensus, off its critical path.

\noindent \textbf{Validator-level load balancing (\S\ref{sec:validator}).}
Each client account is assigned to one validator and submits requests to its workers in a round-robin manner. \sys periodically rebalances assignments by migrating excess load from overloaded validators to those with spare capacity, using load signals derived from the certified DAG. The algorithm runs on every correct validator, and since it operates on data already carried by the DAG, it requires no dedicated exchange of load reports.
\ifextend A migration takes effect only once two records are committed in sequence: its evidence ($2f+1$ matching signed decisions), and then the target validator's acceptance. This committed pair serves as the account's assignment credential from that point on. \sys avoids unnecessary migration when the system is lightly loaded or when no validator has spare capacity to receive load. When the assigned validator is unresponsive, a bounded relay fallback keeps the client account serviceable.\fi

\noindent \textbf{Executor-level load balancing (\S\ref{sec:executor}).}
Within each node, executor workers may be unequally loaded depending on the sharding scheme and workload skewness. \sys schedules transactions across executor workers to balance load, while preserving the consensus-determined order so that execution is serializable and every validator reaches the same state from the same committed sequence, even when validators have different numbers of executor workers~\cite{thomson2012calvin}. Every executor worker receives the full committed batch and independently runs the same lightweight deterministic scheduler, so all workers derive identical assignments without a central dispatcher.
\ifextend Our prototype builds on Narwhal and Tusk~\cite{danezis2022narwhal}, whose scale-out validator architecture provides the certified DAG and multi-worker structure described above; \S\ref{sec:validator} specifies the certified-DAG properties \sys requires.\fi

\section{Validator Level Load Balancing}
\label{sec:validator}

\sys's validator-level load balancing migrates accounts across validators to alleviate overload.
Load balancing in DAG-based BFT protocols is challenging for two main reasons. First, there is no central coordinator: all validators must reach the same migration decision while their local DAG views might diverge. Second, a malicious validator can underreport or overreport its load to mislead the load balancer into incorrect migration decisions, so the protocol must be robust to such misreports.

A configuration epoch $e$ fixes the validator committee, public keys, and load-balancing parameters for its duration. Client accounts are authenticated through a Sybil-resistant mechanism, which also issues each account's initial assignment credential, binding the account to a validator derived deterministically from its identifier. We let $B_{\mathrm{adv}}$ denote an upper bound on the number of admitted accounts controlled by Byzantine clients in one epoch. We consider a fixed epoch throughout the paper and leave reconfiguration out of scope.
\ifextend
The bounded clock-skew assumption introduced below is used only for estimating the magnitude of load-balancing signals and is not required for consensus safety or liveness.
\fi

\begin{figure}
    \centering
    \includegraphics[width=\linewidth]{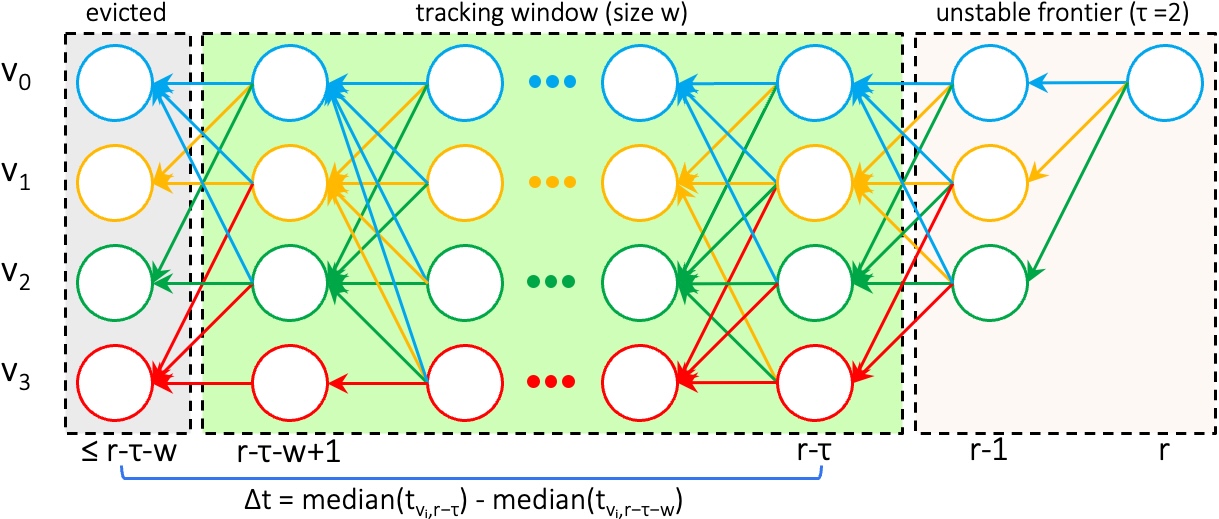}
    \caption{Local DAG view at $v_0$. The active tracking window contains $w$ rounds, from $r{-}\tau{-}w{+}1$ through the stable round $r{-}\tau$ (here $\tau{=}2$)\ifextend; certificates in the unstable frontier are excluded. Each certificate $c_{v_0,r}$ carries the metadata used by Algorithm~\ref{alg:validator}: block proposal timestamp $t_{v_0,r}$, average batch queue delay $\bar{W}(v_0,r)$, and per-account validated transaction counts $n(v_0,r,a)$ for each client account $a$. The evaluation interval $\Delta t = \mathrm{median}(t_{v_i,r-\tau}) - \mathrm{median}(t_{v_i,r-w-\tau})$ anchors the throughput and queue-delay estimates\fi.}
    \label{fig:dag_arch}
\end{figure}

Algorithm~\ref{alg:validator} runs on each validator every $R$ rounds as \emph{evaluation anchor}, concurrently with consensus and off the critical path.
It operates on a stable certified prefix of the DAG: a tracking window of $w$ rounds, from $r{-}\tau{-}w{+}1$ through $r{-}\tau$, where $r$ is the current round and $\tau$ is a configurable value. Excluding the $\tau$ most recent rounds, the \emph{unstable frontier}, lets the window stabilize across validator views. The window size demonstrates a freshness-completeness trade-off: a larger $\tau$ yields more consistent signals at the cost of staler decisions. Figure~\ref{fig:dag_arch} shows a window of size $w$ with $\tau=2$.

Validators run Algorithm~\ref{alg:validator} over their local certified DAG views (not the committed sequence). A natural alternative is to derive migrations deterministically from the committed DAG, which gives all validators identical inputs but delays rebalancing until consensus orders the relevant load history. Certification happens earlier, so using certified metadata keeps load balancing timely. The downside is that honest validators may temporarily hold different certificate sets and thus reach different migration decisions. \sys tolerates this divergence by requiring $2f{+}1$ matching decisions to form migration evidence, so disagreement simply yields no migration rather than an inconsistent assignment. Since only committed migration evidence and receiver acceptance can change an assignment, divergence in the DAG views can delay rebalancing but never compromises assignment safety.

\ifextend
Although we implement \sys on top of Narwhal and Tusk, the design extends to any DAG-based protocols that preserve the certified abstraction, including Bullshark~\cite{spiegelman2022bullshark} and Shoal~\cite{spiegelman2024shoal}. It does not directly carry over to uncertified-DAG protocols such as Mysticeti~\cite{babel2025mysticeti}, which remove the certificate layer that \sys relies on for its load signals and Byzantine-robust aggregation.
\fi

Throughout this section, ties in any deterministic ordering or selection are resolved by identifier (validator or account identity).

\begin{algorithm}[t]
\caption{Validator-level load balancing in \sys}
\label{alg:validator}
\footnotesize
\begin{algorithmic}[1]
\Require (1) per-validator DAG-derived throughput $\tau_v$,
(2) per-account validated transaction counts $n_{v,a}$,
(3) average batch queue delay $\bar{W}_v$,
(4) queue-delay baseline $\tilde{W}_v$,
(5) evaluation interval $\Delta t$,
(6) anchor of the account's latest effective migration $r'_a$, current round $r$,
(7) declared capacity $C_v$,
(8) account-validator latency $L_{v,a}$ (latest committed record),
(9) donor-qualification threshold $W_{\min}$,
(10) migration-cap multiplier $\alpha$,
(11) evaluation interval in rounds $R$

\Ensure Set of migration decisions $\mathcal M=\{(a,v_{\mathrm{old}},v_{\mathrm{new}},r)\}$

\Statex {\color{blue} $\rhd$ Identify overloaded donors and underloaded receivers}
\State $E_v^{s} \gets \max(\tau_v - C_v,\; 0) \quad \forall\, v$ \Comment{structural excess}
\State $\mathcal{D} \gets \{v \mid \bar{W}_v \geq W_{\min} \;\wedge\; E_v^{s} > 0\}$ \Comment{sorted by $(E_v^{s},v)$ desc.}
\State $S_v \gets C_v - \tau_v \quad \forall\, v$ \Comment{declared spare capacity}
\State $\mathcal{R} \gets \{v \notin \mathcal{D} \mid S_v > 0 \;\wedge\; \tau_v > 0\}$
\If{$\mathcal{D} = \emptyset \;\vee\; \mathcal{R} = \emptyset$} \Return $\emptyset$ \EndIf

\State $\mathcal{M} \gets \emptyset$
\ForAll{$d \in \mathcal{D}$}
    \Statex {\color{blue} $\rhd$ Compute migration target}
    \State $\hat{E}_d \gets \tau_d \cdot \frac{\bar{W}_d - \tilde{W}_d}{\Delta t}$ \Comment{inferred excess}
    \State $\gamma_d \gets \big(\tau_d + \max(\hat{E}_d, 0)\big) / \tau_d$ \Comment{incoming-load scale}
    \State $\sigma_d \gets \min\!\big(\max(\hat{E}_d,\; E_d^{s}),\; \alpha \cdot E_d^{s}\big)$ \Comment{migration target}
    \If{$\sigma_d \leq 0$} \textbf{continue} \EndIf

    \Statex {\color{blue} $\rhd$ Select donor's own clients to migrate}
    \State $\mathcal{A}_d \gets \{a \mid n_{d,a} > 0\}$ \Comment{donor's validated clients}
    \State $\rho_a \gets \tau_d \cdot n_{d,a} \;/\; \textstyle\sum_{a' \in \mathcal{A}_d} n_{d,a'} \quad \forall\, a \in \mathcal{A}_d$ \Comment{committed-rate load}
    \State $\ell_a \gets \gamma_d \cdot \rho_a \quad \forall\, a \in \mathcal{A}_d$ \Comment{incoming-rate estimate}
    \State Sort $\mathcal{A}_d$ by $(\rho_a, a)$ descending
    \State $\sigma_{\text{done}} \gets 0$

    \ForAll{$a \in \mathcal{A}_d$}
        \If{$\sigma_{\text{done}} \geq \sigma_d \;\vee\; \rho_a \le 0$} \textbf{break} \EndIf
        \If{$r - r'_a < 2R \;\vee\; \exists\,(a,\cdot,\cdot,\cdot) \in \mathcal{M}$} \textbf{continue} \Comment{cooldown; one decision per account} \EndIf
        \State $v^* \gets \arg\min_{v \in \mathcal{R},\; S_v \geq \rho_a} \big(L_{v,a},\; -S_v,\; v\big)$ \Comment{min latency, max spare, id tie-break}
        \If{$v^* \neq \bot$}
            \State $\mathcal{M} \gets \mathcal{M} \cup \{(a, d, v^*, r)\}$
            \State $\sigma_{\text{done}} \mathrel{{+}{=}} \ell_a;\; S_{v^*} \mathrel{{-}{=}} \rho_a$ \Comment{ reserve at committed rate}
        \EndIf
    \EndFor
\EndFor

\State \Return $\mathcal{M}$
\end{algorithmic}
\end{algorithm}

\subsection{Signal Model and Trust Classification}\label{sec:signal-model}

Every client request carries an account identifier $a$, a nonce $q$ for exactly-once semantics, an application operation, and the identifier of the account's assigned validator, all bound by the client's signature.
If the assigned validator is unresponsive, the client may resubmit in \textsc{relay} mode to a fallback set of $f{+}1$ validators derived deterministically for that account.

We extend the block header of validator $v$ in round $r$ with three fields, covered by the header signature and hence bound by its PoA certificate $c_{v,r}$: the block proposal timestamp $t_{v,r}$, recorded by the proposing primary after its workers' batches are quorum-acknowledged; the average batch queue delay $\bar W_{v,r}$, which measures how long a sealed batch waits before broadcast; and the per-account transaction count $n_{v,r,a}$, the number of \textsc{normal}-mode requests from account $a$ in $v$'s batches of round $r$. Relay copies and requests whose credential names a validator other than~$v$ are excluded from $n_{v,r,a}$, so each request contributes to exactly one account at one validator. Window aggregates discard duplicate $(a,q)$ pairs to prevent replay requests from raising any count. Algorithm~\ref{alg:validator} derives its inputs from these fields, aggregated over the tracking window. The fields differ in how they can be trusted, and we classify them into three signal classes below.

\begin{itemize}[leftmargin=1em,itemsep=1pt,topsep=2pt]
    \item \uline{Validated DAG facts}: the per-account transaction count $n_{v,a}$ records how many transactions from account $a$ were included in validator $v$'s certified batches. These counts are recomputed from certified batch contents. Every validator holding the batches can verify them, and a validator cannot inflate these counts without actually proposing and disseminating the corresponding transactions.
    \item \uline{Clock-derived signals}: we assume bounded clock skew, the clocks of honest validators differ by at most a known bound $\delta$. This assumption is needed neither for the safety nor liveness of the protocol; it affects only the magnitude of the estimates below. The throughput and queue-delay estimates are computed over the evaluation interval $\Delta t$ (Figure~\ref{fig:dag_arch}). The throughput $\tau_v$ is the total certified transaction count across $v$'s batches in the window, divided by $\Delta t$. An evaluation whose $\Delta t$ is not positive is skipped and recurs at the next anchor.
    \item \uline{Self-measured signals}: the average batch queue delay $\bar W_v$ is the mean of the per-round values $\bar W_{v,r}$ over the window, and the baseline $\tilde W_v$ is an exponential moving average of these window means across successive evaluation rounds; it smooths transient spikes. Both are self-measured, but each value is bound by its certified block header, so a validator cannot equivocate about its own reports across peers. These values affect only how much load a validator sheds, never the safety of assignments.
    Two further inputs are fixed in the committed log: the capacity $C_v$, which declares $v$'s safe sustained throughput and is committed once per epoch (we assume an honest validator sets $C_v$ below its saturated throughput, so genuine overload manifests as $\tau_v > C_v$), and the account-validator latency $L_{v,a}$, measured by the client owning $a$ and published as a committed record. Evaluations use the latest $L_{v,a}$ committed before the window opens.
\end{itemize}

\subsection{Load Balancing Algorithm}\label{sec:vlb}

Algorithm~\ref{alg:validator} takes as input the signals defined in \S\ref{sec:signal-model} and outputs a set of migration decisions $\mathcal{M} = \{(a, v_{\mathrm{old}}, v_{\mathrm{new}}, r)\}$, each proposing to move account $a$ from its current validator $v_{\mathrm{old}}$ to a new validator $v_{\mathrm{new}}$ at anchor round $r$. The goal is to identify overloaded validators, select accounts to migrate, and assign them to validators with spare capacity.
These decisions do not change any assignment directly. \ifextend A migration takes effect only when $2f{+}1$ validators produce matching decisions and the receiving validator commits its acceptance. \fi

\noindent \uline{Donors and receivers (lines~1--5).}
A validator is a \emph{donor} (overloaded) when two conditions hold: its certified throughput exceeds its declared capacity, and its average queue delay exceeds a threshold $W_{\min}$ that filters transient noise. The first condition grounds overload in validated data; the second prevents jitter from triggering migration. A non-donor with spare capacity and positive throughput is a \emph{receiver}. The positive-throughput rule excludes idle validators whose spare capacity cannot be verified from their DAG participation alone. If either set is empty, no migration occurs.

\noindent \uline{Migration target (lines~8--11).}
For each donor~$d$, the algorithm estimates how much load to shed using two signals. The \emph{structural excess}~$E_d^s$ captures certified throughput already observed above capacity. The \emph{inferred excess}~$\hat{E}_d$ captures arrival pressure not yet visible in throughput: following the standard fluid view of queues~\cite{kleinrock1974queueing}, sustained queue growth reflects arrivals exceeding service, so the per-window growth rate scaled by throughput estimates the surplus. This is an estimate, since a saturated validator cannot expose its true arrival rate through certified throughput alone.
The migration target takes the larger signal and caps the result at $\alpha$ times the structural excess, where $\alpha \geq 1$ controls aggressiveness. The cap ties migration to validated data: even if a donor inflates its self-reported queue delay, it cannot force migration beyond what its certified throughput justifies.

\noindent \uline{Account selection and receiver assignment (lines~12--23).}
Candidates are the donor's accounts in the tracking window, taken heaviest first (lines~12--15). Each account has two load estimates, one for each side of the migration. The \emph{committed rate}~$\rho_a$ splits the donor's certified throughput by the account's share of validated transactions; since it is fully verifiable, it is what the receiver reserves. The \emph{incoming-rate estimate}~$\ell_a = \gamma_d \cdot \rho_a$ scales $\rho_a$ up because a congested donor's committed rate understates its arrivals; it is what the donor credits toward its target~$\sigma_d$ (line~23). Since $\gamma_d \geq 1$, each account contributes at least $\rho_a$ toward $\sigma_d$, favoring conservative shedding; any residual overload is reconsidered at the next anchor.
An account is skipped if its assignment changed within the last $2R$ rounds, which prevents rapid back-and-forth migration, or if it already has a decision in this invocation (line~19). Otherwise, it goes to the receiver with the lowest latency~$L_{v,a}$ among those with at least $\rho_a$ spare capacity, and is skipped if none qualifies.

The donor classifier uses queue delay rather than raw batch latency because queue delay is the component that grows under sustained overload, regardless of the root cause\ifextend{} (e.g., excess client submissions, outbound broadcast contention, or slow peer acknowledgments)\fi. The correct action is the same: reduce the donor's load by migrating accounts away.

Stratus~\cite{gai2023scaling} uses an analogous end-to-end \emph{stable time} signal for a related purpose; \sys instead decomposes batch confirmation time into queue delay (the overload-sensitive component) and quorum latency (which varies substantially under network imbalance and bandwidth heterogeneity, regardless of overload). Splitting the signals lets us isolate the overload component cleanly and avoid misclassifying network artifacts as overload.

\noindent \textbf{Robustness to misreporting.}
A Byzantine validator~$v$ can misreport its queue delay~$\bar{W}_v$ or capacity~$C_v$. The algorithm limits the damage in each case.
If $v$ inflates $C_v$ to appear as a receiver and later fails to accept, the migration protocol (\S\ref{sec:migration}) protects the affected accounts: an assignment changes only after the receiver's own acceptance is committed, and if the receiver never accepts, the evidence expires and $v$ is excluded from receiver selection for that account for the rest of the epoch.
If $v$ deflates $C_v$ or inflates $\bar{W}_v$ to appear as a donor, the migration target is capped at $\alpha \cdot E_v^s$, which depends on validated throughput and the epoch-committed capacity. Even with $C_v = 0$, the migratable load is bounded by $\tau_v$, the traffic $v$ actually disseminated. The counting rule also limits candidates to $v$'s own accounts, so a faulty donor cannot redirect another validator's clients. An inflated queue delay raises both $\hat{E}_d$ and $\gamma_d$. The first is capped at $\alpha \cdot E^s_d$. The second makes each migrated account count more toward the donor's own target, so the donor reaches its target after shedding fewer accounts. In both cases, the load placed on receivers stays within what they declared.

\ifextend A genuinely slow validator may produce few certificates, making its computed spare capacity misleadingly large. An idle honest validator has $\tau_v = 0$ and is excluded by the receiver rule even though it could be a good receiver. The receiver test cannot distinguish these two cases. \sys bounds the damage through the acceptance requirement and relay fallback. A preventive receiver-health test, such as requiring a minimum certificate count in the window, is compatible with our design; we leave it to future work. \fi

\subsection{Migration Protocol}\label{sec:migration}

Algorithm~\ref{alg:migration} summarizes the migration lifecycle: (1) decision signing, (2) evidence formation, (3) migration activation, and (4) relay fallback. We discuss each step and state the correctness properties.

\begin{algorithm}[t]
\caption{Migration lifecycle at an honest validator $u$}
\label{alg:migration}
\footnotesize
\begin{algorithmic}[1]
\Statex {\color{blue} $\rhd$ At evaluation anchor $r = kR$}
\ForAll{decisions $(a,v_{\mathrm{old}},v_{\mathrm{new}},r)$ output by Alg.~\ref{alg:validator}}
    \State send $m_u = \langle \textsf{\footnotesize MIGRATE},a,v_{\mathrm{old}},v_{\mathrm{new}},r \rangle_{\sigma_u}$ \Comment{at most one per $(a,r)$}
\EndFor

\Statex {\color{blue} $\rhd$ Evidence and activation ($\Pi^+$)}
\State \textbf{upon} collecting $2f{+}1$ matching decisions $\Pi$: submit as \textsc{evidence} record %
\State \textbf{upon} first commit of $\Pi$ at round $r_c$: mark $\Pi$ \emph{live} until activated or until committed round $r_c + T_\Pi$
\Statex \hspace{\algorithmicindent} $\Pi$ is \emph{eligible} while $v_{\mathrm{old}} = v_a \,\wedge\, r > r'_a$
\If{$\Pi$ is live and eligible \textbf{and} $v_{\mathrm{new}} = u$}
    \State submit $\mathsf{acc} = \langle \textsf{\footnotesize ACCEPT},a,v_{\mathrm{old}},v_{\mathrm{new}},r,D(\Pi) \rangle_{\sigma_{v_{\mathrm{new}}}}$
\EndIf
\State let $\Pi^+ = (\Pi,\; \mathsf{acc})$ \Comment{evidence + receiver acceptance}
\State \textbf{upon} commit of $\mathsf{acc}$ while $\Pi$ live and eligible: $A(a) \gets (v_{\mathrm{new}},r)$ %
\State \textbf{upon} expiry of $\Pi$ unactivated: void $\Pi$ and exclude $v_{\mathrm{new}}$ from receiver selection for $a$ for the rest of the epoch

\Statex {\color{blue} $\rhd$ Relay fallback} %
\State \textbf{upon} relay copy $(a,q)$: admit iff
\Statex \hspace{\algorithmicindent} (1) $u \in F(a,q)$, 
\Statex \hspace{\algorithmicindent} (2) client signature and credential valid,
\Statex \hspace{\algorithmicindent} (3) credential matches $A(a)$,
\Statex \hspace{\algorithmicindent} (4) $q$ is the next uncommitted nonce,
\Statex \hspace{\algorithmicindent} (5) no pending fallback copy for $a$ at $u$,
\Statex \hspace{\algorithmicindent} (6) no \emph{effective} fallback request from $a$ in the preceding $R_f$ committed rounds
\State \textbf{upon} admission: reserve $a$ locally until nonce $q$ commits \Comment{committed copy leaves $A(a)$ unchanged, excluded from load counters}
\end{algorithmic}
\end{algorithm}

We write $A(a) = (v_a, r'_a)$ for account $a$'s current assignment and the anchor round of its latest effective migration, initially $(v_0, 0)$, where $v_0$ is the validator derived deterministically from the account identifier during Sybil-resistant admission. A committed object is \emph{effective} if it changes protocol state at its position in the committed log: an effective migration updates $A(a)$, and an effective fallback request is the first valid committed copy of a request, the one that exactly-once semantics executes. Any other committed copy is a \emph{no-op} that changes no state.

\noindent \uline{(1) Decision signing (lines~1--2).}
At each anchor round $r = kR$, a validator running Algorithm~\ref{alg:validator} may sign a migration decision $m_i = \langle \textsf{\footnotesize MIGRATE}, a, v_{\mathrm{old}}, v_{\mathrm{new}}, r \rangle_{\sigma_i}$. An honest validator signs at most one decision per $(a, r)$. Since $R$, $w$, and $\tau$ are fixed within the epoch, the anchor $r$ uniquely identifies the tracking window $\mathcal{W}_r = \{r - \tau - w + 1, \ldots, r - \tau\}$.

\noindent \uline{(2) Evidence formation (lines~3--4).}
A set $\Pi$ of $2f+1$ matching decisions from distinct validators forms \emph{migration evidence}, submitted to the committed log. Once committed at round $r_c$, the evidence is \emph{live} until activated or until $T_\Pi$ committed rounds pass, and \emph{eligible} only while the account is still assigned to $v_{\mathrm{old}}$. The expiry bound $T_\Pi$ prevents stale evidence from lingering indefinitely.

\noindent \uline{(3) Migration activation (lines~5--9).}
An assignment changes only with the receiver's participation. When an honest validator observes committed, live, eligible evidence $\Pi$ naming it as $v_{\mathrm{new}}$, it submits an acceptance record $\mathsf{acc}$ (line~6); its digest $D(\Pi)$ binds $\mathsf{acc}$ to this specific evidence and prevents replays.
Once $\mathsf{acc}$ commits while $\Pi$ is still live and eligible, every honest validator applies $A(a) \gets (v_{\mathrm{new}}, r)$, and $\Pi^+ = (\Pi, \mathsf{acc})$ becomes the account's new assignment credential. No client signature is required, so a client cannot veto a migration. Until activation, the account stays with $v_{\mathrm{old}}$; a request that still reaches $v_{\mathrm{old}}$ afterwards is a no-op and is resubmitted under $\Pi^+$. If the receiver stays silent, $\Pi$ expires after $T_\Pi$ committed rounds, and every honest validator excludes $v_{\mathrm{new}}$ from receiver selection for $a$ for the rest of the epoch.

\ifextend
Two safeguards bound the client's exposure when its assignment changes without its participation: the client's committed latency records constrain where it can be placed, and the installed receiver was vouched for by at least $f+1$ honest validators who evaluated validated counts and committed capacities.
\fi

\noindent \uline{(4) Relay fallback (lines~10--11).}
The assigned validator is the normal route, not the only admission point. If the client receives no reply before its timer expires, it resubmits the same request in \textsc{relay} mode to its fallback set $F(a,q)$ of $f{+}1$ validators, derived deterministically for the account (\S\ref{sec:signal-model}) so that at least one member is honest. An honest member admits the copy only under conditions (1) through (6) of Algorithm~\ref{alg:migration} at line~10.
Two of them bound the fallback traffic: an honest relay holds at most one pending copy per account (5), and an account has at most one effective fallback request every $R_f$ committed rounds (6). 
The committed copy is ordered and executed like any other request but does not change the assignment and is excluded from load counters. 
Several members of $F(a,q)$ may commit copies of the same request; exactly-once semantics on the nonce makes only the first valid committed copy effective, and every later copy is a no-op.

\ifextend
A Byzantine client may invoke the fallback without contacting its assignee. Honest validators reject copies from outside $F(a,q)$, and after commitment, exactly-once semantics and the $R_f$ gate reject further copies. Condition (6) counts only effective requests: committed duplicates that are no-ops neither reset the gate nor block future admission, so replaying past requests cannot censor a client's fallback path.
\fi

\noindent \textbf{Correctness.} We briefly state the correctness of our migration protocol using three properties.

\noindent \uline{P1: Assignment agreement.} For any committed-log prefix, honest validators derive the same effective assignment for every account.

The committed admission record fixes the same initial assignment for each account. Every subsequent transition is triggered by a committed object: an \textsc{evidence} record, an acceptance record, or an expiry deadline measured in committed rounds. Two honest validators that have processed the same committed prefix observe the same objects at the same positions and evaluate the same deterministic conditions, so they apply the same transition, or both treat the object as a no-op. Induction over the log gives agreement.

\noindent \uline{P2: Liveness.} If a correct client submits a valid request with its next nonce and a current assignment credential, the request eventually takes effect.

The underlying consensus protocol provides eventual reliable communication between a correct client and honest validators, and ensures that a valid request submitted by an honest validator is eventually committed. \sys changes only how requests are admitted: a normal request enters through the assigned validator, and a relay copy enters through $F(a,q)$ under conditions (1) through (6). It therefore suffices to show that an honest validator eventually admits the request.

If the assignee is honest, it admits and submits the request, and base-protocol liveness commits it. Otherwise, $F(a,q)$ contains an honest validator, and eventual reliability delivers the relay copy to it. If that validator already holds a pending fallback copy for the account, that copy is eventually committed by base-protocol liveness. Otherwise, once the previous effective fallback request leaves the $R_f$ window, conditions (1) through (6) are satisfied and the copy is admitted; the gate may delay fallback service by up to $R_f$ committed rounds per request but does not permanently block it. \ifextend Since only effective fallback requests occupy the gate, replayed no-ops cannot postpone reopening indefinitely. \fi The base protocol eventually commits at least one admitted copy; the first valid committed copy takes effect and later duplicates are no-ops.

\noindent \uline{P3: Bounded fallback admission.}
The fallback path cannot inject unbounded extra traffic: each request yields at most $f{+}1$ valid copies, at most one committed copy is effective, and an account's effective fallback requests are at least $R_f$ committed rounds apart.

Only the $f{+}1$ members of $F(a,q)$ can submit valid fallback copies of a request, and exactly-once semantics makes at most one committed copy effective. Each honest relay admits at most one pending copy per account, and the $R_f$ gate permits at most one effective fallback request per account per interval. Under the admission bound $B_{\mathrm{adv}}$, honest relay validators admit at most $(f{+}1)B_{\mathrm{adv}}$ fallback copies from Byzantine-client accounts per $R_f$-round interval. Fallback traffic from correct clients with a faulty assignee is their ordinary workload: each request produces at most $f{+}1$ relay copies, rate-limited by the same gate. \ifextend Byzantine validators may inject additional copies into their own batches; such copies lie outside the honest-admission bound. \fi

\ifextend
\paragraph{Censorship by the assignee.} A Byzantine assignee may silently throttle its clients. Affected clients can still submit requests through the relay fallback, but only at the rate the $R_f$ gate allows, and their zero validated counts exclude them from every candidate set in Algorithm~\ref{alg:validator}. However, each effective relay request is a committed fact visible to all honest validators. If an account accumulates enough consecutive effective relay requests in the committed log, this pattern signals that its assignee is not serving it. A deterministic threshold on this count, evaluated identically by all honest validators, could trigger a forced migration through the same evidence-and-acceptance lifecycle, reassigning the account without relying on Algorithm~\ref{alg:validator}'s candidate selection. The client can also publish a high latency record $L_{v,a}$ for the censoring validator, steering future placement away from it at epoch reconfiguration. We leave both extensions to future work.
\fi
\section{Executor Level Load Balancing}
\label{sec:executor}

The previous section focused on load balancing at the validator level. This section addresses load balancing at the executor level, where a committed batch from a validator must be distributed among $m$ executor workers, each managing a distinct shard of the state. While scaling the execution layer by distributing execution responsibilities across workers can enhance performance, it presents two main challenges. First, if the transaction access pattern is skewed, workers may end up processing unequal numbers of transactions, negatively affecting the overall performance.
Second, processing distributed transactions that access records across multiple workers can be challenging. Moreover, adding more workers tends to increase the proportion of distributed transactions due to the finer data sharding.

\sys adopts a \textit{data fusion} model~\cite{lin2016towards}: each transaction is routed to a designated master executor worker that pulls any required remote inputs, executes locally, and installs all writes in its own store without writing state back to the originating executor workers. This reduces the cost of a distributed transaction from a round trip to a single one-way transfer, and subsequent transactions assigned to the same master that access recently fused state incur no transfer at all. We assume read/write sets are statically extractable prior to execution, which holds for Move-based transaction formats. A replicated BFT executor must execute transactions in the consensus-determined order, and nodes may operate different numbers of executor workers, so the scheduler cannot reorder transactions within a batch to improve locality.

A centralized scheduler would become a bottleneck and face the difficulty of synchronizing partition state across workers. Instead, when a batch is committed, each validator worker broadcasts the \textit{full batch} to all $m$ executor workers, and each executor worker independently runs Algorithm~\ref{alg:batch-aware-executor-assignment} on the full batch, then executes its assigned transactions in the consensus-determined order.

This design requires each executor worker to ingest the entire committed stream, covering the records owned by all executor workers, rather than a $b/m$ sub-batch; relative to a sub-batch design that delivers $O(b/m)$ data per executor worker, per-worker LAN ingress grows by a factor of $m$. This in-cast is affordable because the validator workers, primary, and executor workers are mutually trusted components within a single node, typically co-located within the same physical node or the same datacenter with high-bandwidth LAN interconnects (e.g., 10 Gbps), whereas a validator worker's throughput is already capped by the lower WAN egress bandwidth used for inter-validator dissemination. The full-batch broadcast is therefore absorbed whenever the per-executor LAN ingress bandwidth exceeds the worker WAN egress bandwidth by at least a factor of $m$. Since intra-datacenter LAN bandwidth is typically at least an order of magnitude higher than WAN bandwidth, this condition holds for the executor scales we target ($m \leq 10$).

Algorithm~\ref{alg:batch-aware-executor-assignment} presents the execution load balancing in \sys.
The algorithm processes transactions sequentially in the consensus-determined order. For each transaction, it tallies votes from state ownership (lines 4--6): each key in the read/write set votes for the executor worker that will own it after previous transactions are executed. The executor worker is then selected in one of the following ways.
\begin{enumerate}[leftmargin=1em,itemsep=1pt,topsep=2pt]
    \item \uline{Fast path (lines 8--9):} If the entire read/write set maps to one executor worker with available capacity, assign the transaction to that executor worker (zero remote reads).
    \item \uline{Affinity path (lines 10--11):} Otherwise, choose the highest-vote executor worker with sufficient capacity, breaking ties by load, then by ID for determinism.
    \item \uline{Fallback (lines 12--13):} If all candidates are full, assign to the least-loaded executor worker. The per-executor capacity (line 1) is bounded by $\left\lceil \frac{|b|}{m}(1+\epsilon) \right\rceil$ where $\epsilon$ is a configurable load imbalance tolerance parameter.
\end{enumerate}

After each assignment, the algorithm updates load counters and rewrites the keys in a batch-local overlay (lines 14--16). Subsequent transactions compute votes against this overlay, ensuring that scheduling decisions are consistent with the evolving execution plan. This prefix-consistent view is also required by executor workers to determine whether to serve or fetch the state. 

The overlay can be viewed as a sequence of delta updates over the base partition map, which provides a compact representation of the evolving ownership state. This enables re-execution by reconstructing the intermediate ownership state from the base map and the sequence of assignments, without materializing a full updated partition. Between batches, the final overlay $U$ captures the up-to-date data ownership: it serves as the base partition map for scheduling the next batch, and executor workers use it to verify the data ownership of each subsequent transaction's read/write set.

\begin{algorithm}[t]
\caption{Load-Aware Transaction Scheduling in \sys}
\label{alg:batch-aware-executor-assignment}
   \footnotesize
\begin{algorithmic}[1]
\Require batch $b$, executors $m$, partition map $P$, imbalance tolerance $\epsilon$
\Ensure assignment list $\mathcal{A}$ of $(tx, e)$ pairs

\State $cap \gets \lceil \frac{|b|}{m} \cdot (1+\epsilon) \rceil$ \Comment{per-executor capacity}
\State $U \gets P$;\quad $\mathit{load}[1..m] \gets 0$;\quad $\mathcal{A} \gets [\,]$ \Comment{$U$: mutable overlay on $P$}

\ForAll{$tx \in b$} \Comment{in consensus order}
    \Statex {\color{blue} $\rhd$ Tally ownership votes over the overlay}
    \State $votes \gets \emptyset$
    \ForAll{$k \in \textsc{RW}(tx)$}
        \State $votes[U[k]] \mathrel{{+}{=}} 1$
    \EndFor

    \Statex {\color{blue} $\rhd$ Select an executor worker}
    \State $\mathcal{C} \gets \{e : votes[e] > 0 \;\wedge\; \mathit{load}[e] < cap\}$ %
    \If{$|votes| = 1 \;\wedge\; \mathcal{C} \neq \emptyset$}
        \State $e^* \gets$ the element of $\mathcal{C}$ \Comment{fast path: zero remote reads}
    \ElsIf{$\mathcal{C} \neq \emptyset$}
        \State $e^* \gets \arg\max_{e \in \mathcal{C}} \big(votes[e],\ -\mathit{load}[e]\big)$ \Comment{affinity path}%
    \Else
        \State $e^* \gets \arg\min_{e} \mathit{load}[e]$ \Comment{fallback: all candidates at capacity}
    \EndIf

    \Statex {\color{blue} $\rhd$ Record the assignment and update the overlay}
    \State append $(tx, e^*)$ to $\mathcal{A}$;\quad $\mathit{load}[e^*] \mathrel{{+}{=}} 1$
    \ForAll{$k \in \textsc{RW}(tx)$}
        \State $U[k] \gets e^*$
    \EndFor
\EndFor
\State \Return $\mathcal{A}$
\end{algorithmic}
\end{algorithm}

The algorithm is order-preserving and deterministic: it processes transactions in consensus-determined order, uses deterministic tie-breaking, and derives all state from the batch and initial partition map. Executor workers with the same inputs therefore produce identical assignments without a central dispatcher, and nodes agree on execution regardless of their worker count.

\section{Evaluation}
\label{sec:eval}

The goal of our evaluation is to quantify how effectively \sys mitigates the performance degradation caused by load imbalance in DAG-based BFT protocols.
\sys performs transaction submission routing and execution scheduling at the ordering and execution layer respectively. We evaluate each layer independently because the two are subject to different bottlenecks and thus require different baselines. At the ordering layer, we compare \sys against vanilla Narwhal~\cite{danezis2022narwhal} to quantify both the performance gain under skewed workloads and the overhead under balanced workloads. At the execution layer, the closest scale-out baseline is PilotFish~\cite{kniep2025pilotfish}, which fixes object-to-worker placement at dispatch time and thus offers no load-aware scheduling. We separately measure (1) throughput of \sys's executor with load-aware scheduling disabled, evaluated by replaying committed batches, and (2) the isolated scheduling cost versus remote-read trade-off against Hermes-style prescient routing.

At the ordering layer, \sys augments Narwhal block headers, and hence the certificates over them, with additional metadata: per-account validated transaction counts, the block proposal timestamp, and the average batch queue delay. The load signals therefore piggyback on the DAG, so a validator computes and verifies its decisions from certificates with no dedicated metric-exchange round. \ifextend The evaluated prototype implements this signal-and-decision path, which \S\ref{sec:eval-misreport} additionally exercises under validators that misreport their self-declared signal; the migration-commit lifecycle remains a protocol-level Byzantine-safety mechanism whose guarantees are established analytically in \S\ref{sec:migration}. \fi
At the execution layer, \sys runs the full-batch decentralized scheduling design of \S\ref{sec:executor}.

\subsection{Experimental Setup}
\label{sec:eval-setup}

Experiments were conducted on CloudLab~\cite{duplyakin2019design} using \texttt{xl170} nodes\ifextend, each equipped with an Intel Xeon E5-2640 v4 (10 physical cores, 20 hardware threads), $64$ GB DDR4 ECC memory, and a $480$ GB SSD, connected via up to $25$ Gbps links\fi. WAN conditions (bandwidth and $100$ ms RTT) are shaped by the CloudLab profile. Within each validator, we cap bandwidth at $300$ Mbps for the primary and $600$ Mbps for the worker to mirror a scaled-out DAG-based deployment, where operators prioritize bandwidth for consensus traffic. \ifextend System components are pinned to disjoint CPU sets at the hardware-thread level: $10$ logical CPUs for workers, $6$ for the primary, and $2$ for clients, with the remainder reserved for the OS. Worker batch storage uses an in-memory HashMap-backed store to isolate network and consensus bottlenecks from disk I/O, which is orthogonal to our load-balancing contribution; the primary's storage remains RocksDB-backed.\fi

We use the {\sf SmallBank} benchmark with $1$ M users. Each validator is paired with one benchmark client that simulates a geo-distributed region. Account identifiers are partitioned into equal contiguous ranges, and the identifier-derived initial assignment (\S\ref{sec:migration}) maps each range to the region's co-located validator; each client submits transactions on behalf of its region's accounts at a fixed rate. The user for each transaction is selected according to the configured skewness, and transactions are padded with zeros up to the configured size. \ifextend To prevent TCP backpressure from both throttling the client sending rate and inflating tail latency, the client buffers requests in an unbounded channel.\fi

The baseline validator configuration uses $4$ validators each running a single worker, a $500$ KB batch size, a $1$ KB block size, $200$ ms maximum batch and header delay, and a transaction size of $512$ B. The load balancing algorithm (Algorithm~\ref{alg:validator}) is invoked every $R{=}60$ rounds over a $w{=}30$-round tracking window. We set the migration-amount cap $\alpha = 4$; larger values shed more load per invocation but risk ping-pong oscillation between donor and receiver. We set the donor queue-delay threshold $W_{\text{min}} = 30$~ms to filter out network jitter\ifextend{; $W_{\text{min}}$ is the minimum queue-delay threshold a validator must exceed to be classified as a donor}\fi. Each validator computes its capacity $C_v$ (\S\ref{sec:validator}) from its egress bandwidth, since broadcast fan-out saturates the network before CPU or disk in a scaled-out DAG-based BFT deployment.
\ifextend
Concretely, $C_v$ is the sustainable egress throughput of the validator's broadcast:
\begin{equation}
\label{eq:capacity}
C_v = \left\lfloor \frac{BW}{(s + 44)\,(n-1)} \cdot 0.8 \right\rfloor,
\end{equation}
where $BW$ is the validator's egress bandwidth, $s$ is the transaction payload size in bytes, and the constant $44$ bytes accounts for per-transaction wire overhead ($40$\,B TCP/IP header plus $4$\,B length-prefix codec overhead). The $(n{-}1)$ factor reflects that each worker broadcasts every transaction to all other $n{-}1$ validators, and the $0.8$ multiplier leaves headroom for control traffic and short bursts. Other components (e.g., primary CPU or storage) may become limiting in setups with narrower WAN links, and $C_v$ can be adjusted accordingly.

The per-account-per-validator RTT $L_{v,a}$, used by the migration policy to select a receiving validator for a shed account, is $100$ ms for remote-region submissions and $0$ ms locally. We use a garbage collection (GC) depth of $600$ rounds to prevent blocks and certificates needed by slow validators from being garbage collected before they catch up.
\fi
Each run lasts 15 minutes unless stated otherwise. Clients first submit a balanced workload for 4 minutes (the \emph{balanced phase}) and then switch to the configured skewness for the remainder (the \emph{imbalanced phase}).  \ifextend Committed throughput over time is computed as the total number of committed transactions across validators binned at $10$ seconds granularity, smoothed with a 3-bin moving average. Latency is measured as the interval between transaction submission and the time at which the transaction is committed by the consensus protocol, tracked via sample transactions under high load.\fi

\begin{figure*}[htbp]
    \centering
    \begin{subfigure}{0.24\textwidth}
        \centering
        \includegraphics[width=\linewidth]{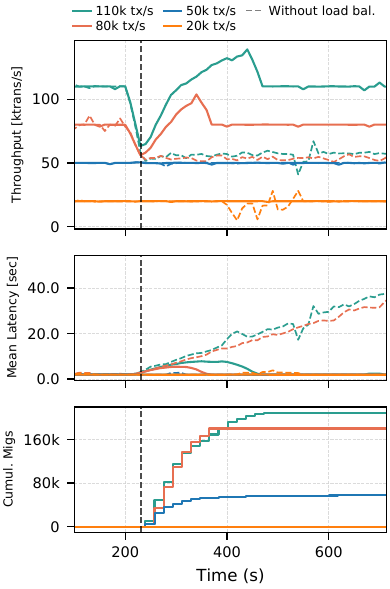}
        \caption{Impact of Submission rate \quad ($n=4$, $90\%$ skewness)}
        \label{fig:config_a}
    \end{subfigure}\hfill
    \begin{subfigure}{0.24\textwidth}
        \centering
        \includegraphics[width=\linewidth]{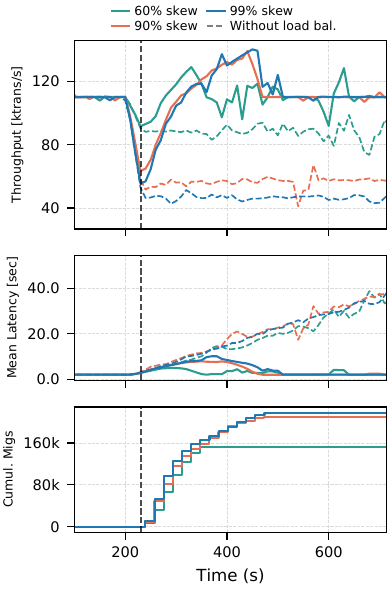}
        \caption{Impact of Load Skewness \qquad ($n=4$, $110$ ktps)}
        \label{fig:config_b}
    \end{subfigure}\hfill
    \begin{subfigure}{0.24\textwidth}
        \centering
        \includegraphics[width=\linewidth]{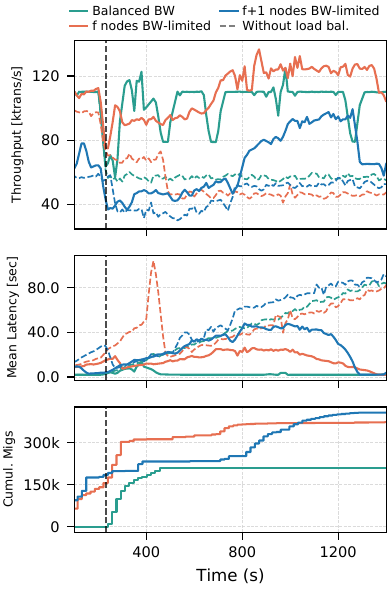}
        \caption{Impact of Bandwidth ($n=4$, $90\%$ skewness, $110|65$ ktps, extended)}
        \label{fig:config_c}
    \end{subfigure}\hfill
    \begin{subfigure}{0.24\textwidth}
        \centering
        \includegraphics[width=\linewidth]{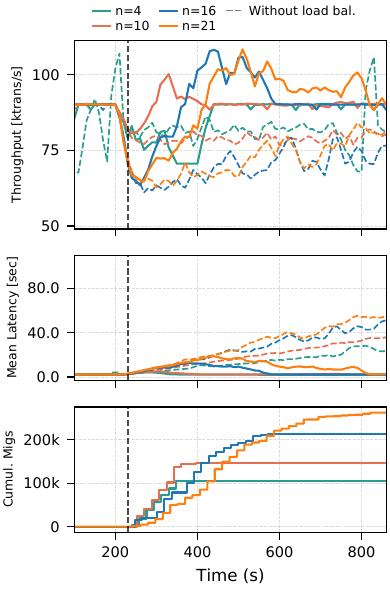}
        \caption{Impact of Scalability ($60\%$ skewness, $90$ ktps)}
        \label{fig:config_d}
    \end{subfigure}
    \caption{Validator level load balancing evaluation}
    \label{fig:validator_eval}
\end{figure*}

\subsection{Validator Level Load Balancing}

We evaluate validator level load balancing along six axes\ifextend{}: submission rate (\S\ref{sec:eval-submission-rate}), workload skewness (\S\ref{sec:eval-skewness}), per-validator bandwidth heterogeneity (\S\ref{sec:eval-bandwidth}), number of validators (\S\ref{sec:eval-scale}), adaptivity to shifting hotspots (\S\ref{sec:eval-adaptivity}), and robustness to misreporting validators (\S\ref{sec:eval-misreport})\fi. Our goal is to demonstrate that \sys (1)~rebalances skewed workloads to recover throughput and latency, (2)~distributes the load efficiently across heterogeneous validators,
(3)~introduces negligible overhead when no rebalancing is required, (4)~scales with the number of validators, and (5)~remains effective when validators misreport their self-declared load signal.

We define workload imbalance as $60\%$ or $90\%$ of the total load being directed at $25\%$ of the validators, and validator heterogeneity as up to $f$ or $f{+}1$ validators being bandwidth-capped at $200$ Mbps. The highest submission rate we evaluated, $110$ ktps, is roughly $76\%$ of the aggregate service throughput, which allows us to observe whether \sys can dissipate the backlog built up during the imbalanced phase in a reasonable time.

\subsubsection{Impact of Submission Rate}
\label{sec:eval-submission-rate}

We evaluate \sys under four submission rates: $20$, $50$, $80$ and $110$ ktps, with the imbalanced phase skewness fixed at $90\%$.
Figure~\ref{fig:config_a} shows throughput (top), mean latency (middle), and cumulative account migrations (bottom).
During the balanced phase, \sys matches the baseline on both throughput and latency, confirming that the load-balancing algorithm imposes negligible overhead when no migration is required.
The brief dip and rebound in the 110 ktps run before the phase change is a consensus-ordering artifact in which one validator's certified batches are temporarily not referenced by leader blocks and are then ordered together once referenced; certification throughput remains flat over this window and no migrations are triggered.

Under a light workload of $20$ ktps, even routing $90\%$ of the load to a single validator remains well below its processing capacity, and \sys correctly migrates no accounts.
At higher rates, the onset of the imbalanced phase reduces performance for both \sys and the baseline (which does not employ load balancing techniques). \sys, however, dissipates the backlog and recovers throughput and latency, whereas the baseline does not.
At $50$ ktps, \sys operates near the saturation knee of the first validator: an initial burst of migrations rebalances the load, after which no further rebalancing is induced. \ifextend During the initial phase, the queue-delay differential $(\bar{W}_v - \tilde{W}_v)$ is positive, causing Algorithm~\ref{alg:validator} to classify $v_0$ as a donor; once the system stabilizes, the delay difference becomes small or negative. \fi At $80$ and $110$ ktps, both $E_d^s$ and $\hat{E}_d$ are larger, driving more migrations and a longer backlog drain. \ifextend Convergence time grows monotonically with submission rate, as the cumulative migration curves make explicit.\fi

\begin{figure}
    \centering
    \includegraphics[width=\linewidth]{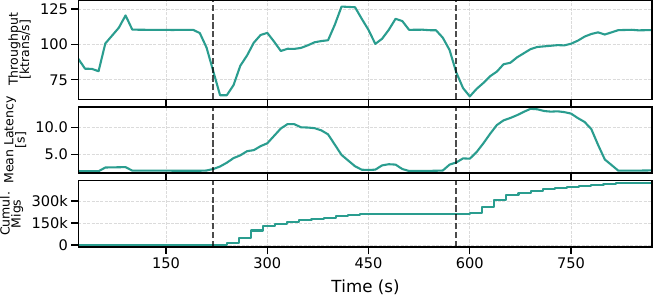}
    \caption{Adaptivity of \sys under a shifting hotspot. \ifextend Balanced phase (0--4\,min); first imbalanced phase with hotspot on $v_0$ (4--10\,min); second imbalanced phase with hotspot shifted to $v_1$ (10--15\,min). \sys issues a second migration burst at the 10-minute shift and recovers throughput in both phases.\fi}
    \label{fig:adaptivity}
\end{figure}

\subsubsection{Impact of Workload Skewness}
\label{sec:eval-skewness}

Figure~\ref{fig:config_b} fixes the submission rate at $110$ ktps and varies skewness from $60\%$ to $90\%$ to $99\%$. Higher skewness concentrates more load on fewer validators, raising the overload on the donor and thus the migration target. As a result, \sys migrates more accounts and takes longer to drain the backlog: the $99\%$-skew run converges roughly two minutes later than the $60\%$-skew run, with the $90\%$-skew case in between.
The cumulative migration curve scales accordingly, and the recovered throughput under all three skewness settings approaches the balanced-phase level, whereas the baseline remains degraded throughout the imbalanced phase.

\subsubsection{Impact of Bandwidth Heterogeneity}
\label{sec:eval-bandwidth}
 
We next evaluate \sys under heterogeneous validator bandwidth. A \emph{slow validator} is one whose per-worker egress bandwidth is capped at $200$ Mbps (instead of the $600$ Mbps default); its primary-side bandwidth is unchanged. This models realistic deployments where a subset of operators runs on degraded WAN links. We evaluate two configurations: up to $f$ slow validators (which the quorum threshold $2f{+}1$ can mask) and $f{+}1$ slow validators (which it cannot). In the first configuration, the submission rate is $110$ ktps, while in the second configuration we limit the submission rate to $65$ ktps. 

Convergence here is paced by the donor condition of Algorithm~\ref{alg:validator}. Queue delay alone never qualifies a validator as a donor: donor qualification and the per-anchor target $\alpha \cdot E_d^s$ are anchored in validated throughput above committed capacity, and a saturated validator's certified throughput understates its arrival rate (\S\ref{sec:validator}), so its structural excess stays small and load is shed through a sequence of bounded migration waves, visible as the staircase in the cumulative-migration curves, with residual overload recurring at later anchors by design. We therefore extend these runs to $25$ minutes, keeping the $4$-minute balanced phase, so the full multi-wave convergence and the post-recovery steady state fall within the measurement window. The results are presented in Figure~\ref{fig:config_c}.

With $f$ bandwidth-limited nodes, \sys begins migrating accounts away from the slow validators immediately, improving throughput during the balanced phase. When the imbalanced phase starts, both \sys and the baseline degrade; a second migration wave from the overloaded $v_0$ restores throughput to near the offered rate, and the backlog then drains gradually through small jitter-triggered residual waves, with mean latency returning to the balanced-phase level in the later part of the run.
 
With $f{+}1$ bandwidth-limited nodes, forming a quorum requires $2f{+}1$ validators and the $(2f{+}1)$-th fastest validator is itself slow, so no migration can restore the balanced-bandwidth commit rate. The slow validators still satisfy the donor condition of Algorithm~\ref{alg:validator}, so \sys sheds accounts toward the fast validators from the start of the run; a second migration wave mid-run sheds further load, after which committed throughput temporarily exceeds the offered $65$ ktps while the backlog drains and mean latency returns near the balanced-phase level by the end, whereas the baseline's latency grows for the remainder of the run.

\subsubsection{Scalability with Number of Validators}
\label{sec:eval-scale}

Figure~\ref{fig:config_d} evaluates \sys across $n \in \{4, 10, 16, 21\}$ validators at a fixed offered rate of $90$ ktps. Unlike the preceding experiments, increasing $n$ reduces each validator's modeled capacity because every worker broadcasts each batch to $n{-}1$ peers\ifextend{} (Eq.~\ref{eq:capacity})\fi. Consequently, $110$ ktps corresponds to approximately $76\%$ of the modeled aggregate capacity at $n{=}4$ but approximately $97\%$ at $n{=}21$, leaving almost no spare capacity for absorbing migrated load and draining the backlog. We therefore fix the rate at $90$ ktps, approximately $80\%$ of the minimum aggregate capacity across the evaluated committee sizes, so that every configuration retains comparable headroom and the observed differences are attributable to committee size rather than saturation.

During the balanced phase, when no migrations are required, \sys's overhead is negligible and its performance matches the baseline across all network sizes. When the imbalanced phase begins, $60\%$ of the load is concentrated on $25\%$ of the validators (i.e., $\{1, 3, 4, 6\}$ donors respectively), and all configurations degrade. \sys recovers in every configuration: throughput briefly exceeds the offered rate while the backlog drains and then settles back at $90$ ktps, and mean latency returns to the balanced-phase level. Recovery takes longer, and cumulative migrations grow, as the network size increases.

\ifextend As $n$ grows, the per-donor skewed load $\tau_d$ shrinks (the total rate is split across more donors, roughly $0.25n$ of them), and the per-donor capacity $C_d$ also shrinks because the broadcast fan-out factor $(n{-}1)$ in Eq.~\ref{eq:capacity} reduces each worker's sustainable throughput. Both terms are $O(1/n)$, so in the overload regime ($\tau_d > C_d$) their difference $E_d^s = \tau_d - C_d$ is $O(1/n)$ as well. The migration cap $\alpha \cdot E_d^s$ shrinks accordingly, so \sys sheds less load per invocation at larger $n$, which explains the longer convergence time and also suppresses ping-pong. \fi

\subsubsection{Adaptivity to Shifting Hotspots}
\label{sec:eval-adaptivity}
 
\ifextend The preceding experiments show that \sys converges once after a single static workload shift. \fi Hotspots can also move over time, and the load balancer must keep up with them rather than lock in the first assignment. To evaluate this, we run a three-phase workload at $110$ ktps and $90\%$ skew on $n{=}4$ validators: a balanced phase (0--4\,min) matching the earlier setup, a first imbalanced phase concentrating load on $v_0$ (4--10\,min), and a second imbalanced phase in which the hotspot shifts to $v_1$ (10--15\,min).
Figure~\ref{fig:adaptivity} shows that \sys migrates accounts away from $v_0$ at the onset of the first shift and recovers throughput, as in \S\ref{sec:eval-submission-rate}. When the hotspot moves to $v_1$ at 10 minutes, the queue-delay signal on $v_1$ exceeds $W_{\text{min}}$, triggering a second migration burst that re-routes accounts toward the now-underloaded $v_0$ and the remaining validators; throughput recovers a second time. The no-load-balancing baseline remains degraded across both imbalanced phases. This confirms that \sys's load balancer responds to ongoing load shifts rather than acting as a one-shot rebalancer.

\begin{figure}[t]
    \centering
    \begin{subfigure}{0.49\linewidth}
        \centering
        \includegraphics[width=\linewidth]{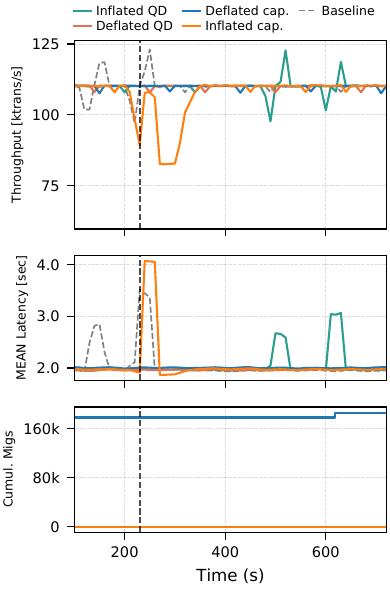}
        \caption{Balanced workload, $110$ ktps}
        \label{fig:misreport-bal}
    \end{subfigure}\hfill
    \begin{subfigure}{0.49\linewidth}
        \centering
        \includegraphics[width=\linewidth]{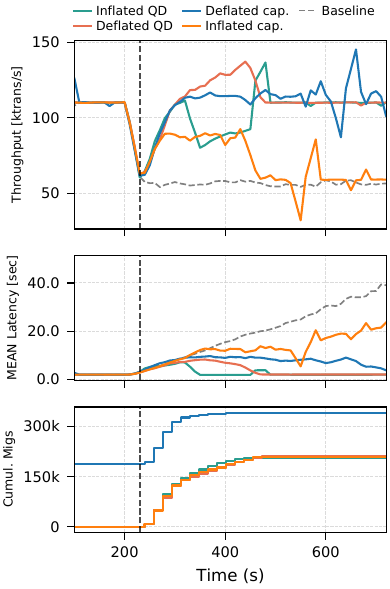}
        \caption{$90\%$ skew, $110$ ktps}
        \label{fig:misreport-skew}
    \end{subfigure}
    \caption{Robustness to misreporting. The last validator falsifies one signal per run. Under a balanced workload (a), no attack degrades performance. Under skew (b), only inflated capacity, the one signal unverifiable at commit time, has an effect: one bounded migration burst and degraded performance for the remainder of the epoch.}
    \label{fig:misreport}
\end{figure}

\subsubsection{Robustness to Misreporting Validators}
\label{sec:eval-misreport}

We stress the two self-reported inputs of Algorithm~\ref{alg:validator} by reusing the $110$ ktps run from \S\ref{sec:eval-submission-rate} and having the last validator falsify one signal per run: the queue delay $\bar W_v$ or the capacity $C_v$, inflated or deflated. Under a balanced workload, no attack degrades performance (Figure~\ref{fig:misreport-bal}), since no validator qualifies as a donor and the misreports have nothing to act on.

Under $90\%$ skew, three of the four attacks still track the honest baseline (Figure~\ref{fig:misreport-skew}). Deflating $\bar W_v$ or $C_v$ only removes the misreporting validator from its own donor or receiver role. Inflating $\bar W_v$ raises the donor's target, but that target is capped at $\alpha \cdot E_d^s$, computed from validated throughput and cannot be inflated by self-reports.

Inflated capacity behaves differently. A capacity declaration is a commitment about future service, and no certified evidence contradicts it at commit time (\S\ref{sec:vlb}). The misreporting validator triggers a single migration burst, still capped per anchor by each donor's $\alpha \cdot E_d^s$, after which the migration curve flattens. Throughput and latency stay degraded for the rest of the run: the inflated capacity hides the validator's structural excess, so it never re-qualifies as a donor within the epoch.

Two safeguards bound the harm. Its magnitude is capped by the donors' per-anchor targets, and its duration is capped by the epoch boundary, where capacity is re-committed. Meanwhile, affected clients keep making progress through the relay fallback. Both mechanisms are specified in \S\ref{sec:migration} and are established analytically.

\subsection{Executor Level Load Balancing}

We evaluate the executor level along two axes: a scalability experiment that replays committed batches on CloudLab to measure throughput, and a scheduler microbenchmark that isolates scheduling cost vs.\ remote-read trade-off against Hermes-style routing~\cite{lin2021don}.

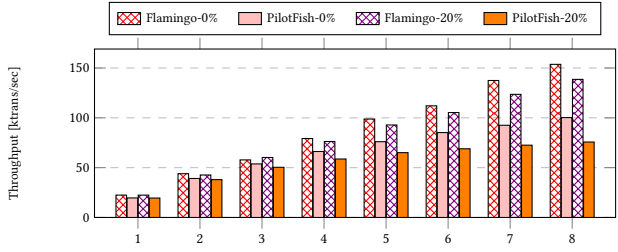
\begin{figure}[t]
\vspace{-1em}
\tiny
\begin{tikzpicture}
    \begin{axis}[
       ybar=0.5*\pgflinewidth,
        bar width=0.14cm,
        width=\columnwidth,
        height=.45\columnwidth,
        ymajorgrids = true,
        grid style=dashed,
        ylabel={Throughput [ktrans/sec]},
        ylabel style={yshift=-1em},
        symbolic x coords={$1$,$2$,$3$,$4$,$5$,$6$,$7$,$8$},
        xtick = data,
        enlarge x limits=0.1,
        ymin=0,
        legend columns=4,
        legend style={
            at={(0.5,1.28)},
            anchor=north,
            font=\tiny,
            /tikz/every even column/.append style={column sep=0.2cm},
        },
        legend image code/.code={%
            \draw[#1] (0cm,-0.08cm) rectangle (0.3cm,0.08cm);%
        },
    ]
     
    \addplot[style={black, pattern color=red,pattern = crosshatch},mark=none]
    coordinates {($1$,22.437)($2$,43.917)($3$,57.764)($4$,79.215)($5$,98.789)($6$,112.033)($7$,137.488)($8$,153.687)};
    \addplot[style={black,fill=pink},mark=none]
    coordinates {($1$,19.590)($2$,39.120)($3$,53.735)($4$,66.191)($5$,76.040)($6$,85.188)($7$,92.533)($8$,100.221)};
    
    \addplot[style={black,pattern color=violet,pattern = crosshatch},mark=none]
    coordinates {($1$,22.412)($2$,42.623)($3$,60.202)($4$,76.262)($5$,92.831)($6$,105.311)($7$,123.572)($8$,138.593)};
    \addplot[style={black,fill=orange},mark=none]
    coordinates {($1$,19.489)($2$,37.946)($3$,50.350)($4$,58.678)($5$,65.040)($6$,68.928)($7$,72.559)($8$,75.720)};
\addlegendentry{\sys-$0\%$}
\addlegendentry{PilotFish-$0\%$}
\addlegendentry{\sys-$20\%$}
\addlegendentry{PilotFish-$20\%$}
    \end{axis}
\end{tikzpicture}
\vspace{-1em}
\caption{Executor throughput under $50\%$ initial skew. \sys vs.\ PilotFish as executor/validator workers scale from $1$ to $8$, for single-shard (\texttt{dist\_tx}$\,{=}\,$0) and $20\%$ distributed (\texttt{dist\_tx}$\,{=}\,$0.2) workloads.}
\label{fig:e-scaliablity}
\vspace{-1em}
\end{figure}

\subsubsection{Scalability Experiment}
\label{sec:eval-exec-scale}

Using the same CloudLab nodes as \S\ref{sec:eval-setup}, we connect each Narwhal worker to its executor workers via $10$ Gbps LAN links, which carries the full-batch broadcast required by our decentralized scheduling design.
Each validator worker commits roughly $25$ ktps.
To stress load imbalance, we skew the initial partition so that $50\%$ of transactions are assigned to the first executor worker.
We vary the number of executor workers (and, proportionally, validator workers) from $1$ to $8$ and measure throughput.
The \texttt{dist\_tx} parameter controls the fraction of \texttt{SendPayment} transactions, which touch two accounts and therefore may become cross-shard; when \texttt{dist\_tx}$\,{=}\,$0 the workload is entirely single-shard.
We tune the executor service throughput to be only slightly above the committed rate so that any imbalance directly hurts throughput.
 
Figure~\ref{fig:e-scaliablity} shows the results.
\sys outperforms PilotFish across all executor worker counts in both workload variants.
Both systems improve with scale-out, but \sys shows a larger gain because its scheduler redistributes load across executor workers.
In the single-shard case (\texttt{dist\_tx}$\,{=}\,$0), throughput scales near-linearly for \sys: adding executor workers increases aggregate service capacity with minimal cross-executor coordination, and \sys reaches $\sim 154$ ktps at $m{=}8$ (about a $6.9\times$ speedup over $m{=}1$), while PilotFish saturated near $100$ ktps.
When $20\%$ of transactions are distributed (\texttt{dist\_tx}$\,{=}\,$0.2), scaling becomes sub-linear because each additional executor worker increases the probability that a distributed transaction's read and write sets span different shards, requiring remote reads and coordination that offset the added parallelism. Nonetheless, \sys sustains a consistent advantage, validating the practicality of full-batch decentralized scheduling.


\begin{figure}[t]
    \centering
    \includegraphics[width=\linewidth]{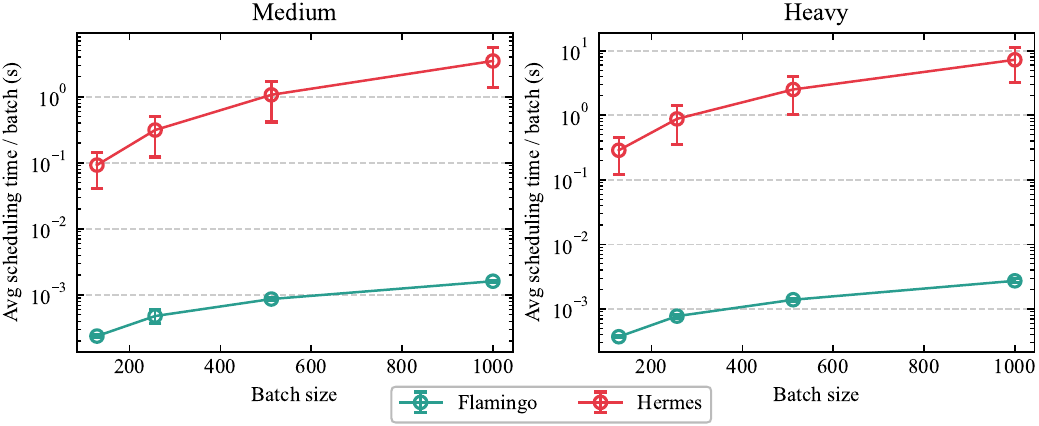}
    \caption{Scheduler runtime per batch as batch size grows (log scale). \sys's single forward pass stays at the millisecond level; Hermes's iterative rerouting reaches multi-sec range.}
    \label{fig:e-scheduling-time-batch}
\end{figure}
 
\subsubsection{Scheduler Microbenchmark}
\label{sec:eval-exec-sched}

SmallBank transactions touch at most two accounts, which is too narrow to exercise scheduling quality meaningfully.
We therefore use a synthetic key-set workload with controllable transaction size and temporal overlap.
Transaction sizes are drawn from three distributions:
\textsc{Light} ($2$ keys with $0.8$ probability, $4$ with $0.2$ probability: $2/0.8$, $4/0.2$),
\textsc{Medium} ($2/0.5$, $4/0.3$, $8/0.2$), and
\textsc{Heavy} ($4/0.5$, $8/0.3$, $16/0.2$).
The \emph{overlap probability} controls how likely a new transaction's accounts are drawn from a recent history window of previously seen accounts; higher values create temporal locality where the same accounts recur in bursts.
We fix $8$ executor workers, $1$ M accounts with range partitioning, batch size $1000$, and report averages over $5$ batches.
 
We compare against Hermes using a scheduler-level simulator.
This choice is deliberate: Hermes's prescient routing is defined over a reordered batch, and its benefit comes from batch-wide lookahead and rerouting.
Reproducing its full-system impact would require fixing to a specific communication model, object size, and migration cost that are orthogonal to the scheduling question.
Our simulator fixes the synthetic workload above and measures total remote reads and scheduler runtime.
\ifextend Following Hermes's simplification, transactions expose explicit read and write sets, and persistent ownership changes are applied through the write set. \fi
 
Figure~\ref{fig:e-scheduling-time-batch} shows scheduling time as batch size grows.
\sys remains at the millisecond level across all batch sizes and transaction types (roughly $2$ ms for a batch size of $1000$), while Hermes grows into the multi-second range; a gap of approximately three orders of magnitude.
This is expected because Hermes optimizes over a permuted batch and performs iterative rerouting to satisfy load constraints, whereas \sys makes a single deterministic forward pass that preserves the agreed-upon transaction order.

\begin{figure}[t]
    \centering
    \includegraphics[width=\linewidth]{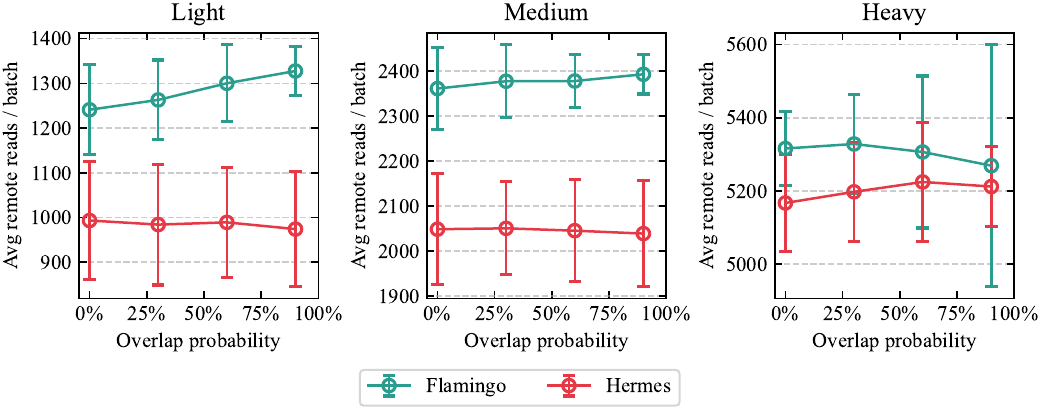}
    \caption{Average remote reads per batch vs.\ overlap probability across transaction sizes. Error bars span $3$ seeds.}
    \label{fig:e-rr-overlap}
\end{figure}

Figure~\ref{fig:e-rr-overlap} shows total remote reads as overlap probability varies.
Hermes achieves fewer remote reads across all transaction sizes because reordering lets it exploit future locality that \sys cannot use. The gap is consistent: roughly $20$--$30\%$ for light and medium transactions, narrowing for heavy transactions where the larger key sets make most transactions inherently cross-shard regardless of scheduling. \ifextend Error bars (3 seeds) confirm that the gap is stable, though variance increases with transaction size. \fi Despite this gap, the trade-off favors \sys: a moderate increase in remote reads buys a $\sim 1000 \times$ reduction in scheduling time.

\ifextend
These results support our design choice: \sys recovers much of the locality benefit of prescient routing while remaining cheap enough to run independently on every executor worker.
\fi

\subsection{End-to-End Evaluation}
\label{sec:eval-e2e}

The preceding sections evaluate the two load-balancing layers in isolation. However, their effects are coupled in a full deployment: validator migration changes where the requests are proposed, and the executor layer must absorb whatever the ordering layer commits. We therefore run the full pipeline, from client submission through consensus to execution, and measure both \emph{committed} performance and \emph{end-to-end} (E2E) performance (transactions committed and executed).

\begin{figure}[t]
    \centering
    \includegraphics[width=\linewidth]{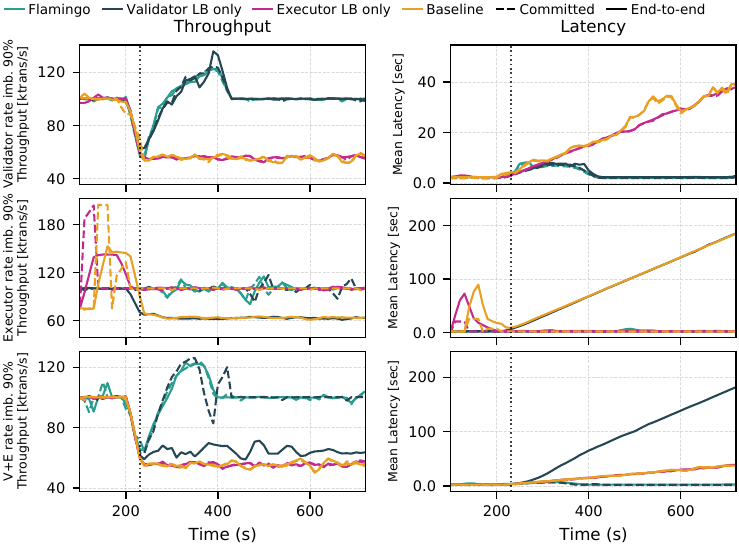}
    \caption{E2E evaluation under validator-level (top), executor-level (middle), and combined (bottom) $90\%$ imbalance. Dashed lines: committed (consensus) performance; solid lines: E2E (committed and executed) performance.}
    \label{fig:e2e}
\end{figure}

We deploy $n{=}4$ validators, each attached to $3$ executor workers, at $110$ ktps rate with the two-phase workload from the experimental setup as in \S\ref{sec:eval-submission-rate}. We induce imbalance at $90\%$ in three ways: (i) \emph{validator rate imbalance}, where $90\%$ of submissions target accounts assigned to a single validator; (ii) \emph{executor rate imbalance}, where $90\%$ of transactions touch accounts residing in a single executor worker's shard; and (iii) both simultaneously. We compare four variants: full \sys, \sys with validator-level load balancing only, \sys with executor-level load balancing only, and the baseline (no load balancing). Figure~\ref{fig:e2e} shows the results.

Under validator rate imbalance (top), the bottleneck lies before commitment, so only the variants with validator-level load balancing recover: full \sys and the validator-only variant dissipate the backlog and return committed and E2E throughput to the offered rate, with latency returning to the balanced-phase level. The executor-only variant tracks the baseline: no scheduling decision made after commitment relieves an overloaded ordering path.

Under executor rate imbalance (middle), it inverts. Consensus is unaffected, and all variants sustain the offered committed rate. Without executor-level load balancing, E2E throughput settles well below the committed rate and E2E latency grows as the execution backlog accumulates. Full \sys and the executor-only variant keep E2E throughput matched to the committed rate with flat latency, confirming that the executor scheduler resolves an imbalance that is invisible at the consensus layer.

The combined imbalance (bottom) shows the bottleneck shifting from one layer to another. Only full \sys restores E2E performance to the offered rate. The validator-only variant resolves the upstream bottleneck but exposes the downstream one: it admits and commits the full rate, but the committed stream lands on the hot executor worker, so a post-commit execution backlog accumulates and its E2E latency grows even though every transaction commits promptly.
The executor-only variant and the baseline remain bottlenecked at the ordering layer throughout.
\ifextend The baseline never reaches the executor bottleneck: its collapsed ordering layer throttles admission to roughly the hot executor's capacity, so its executors keep up and its latency growth stems from pre-commit queueing at the hot validator instead. The plotted mean latency also understates the baseline's degradation: latency samples are drawn uniformly across regions while the baseline's congestion is concentrated in the single hot region, whereas the validator-only variant's executor backlog delays all regions alike. \fi
These results show that the two mechanisms compose without interference and that each layer is necessary for the bottleneck it targets: resolving only one layer moves the bottleneck rather than removing it.
\section{Related Work}\label{sec:related}

\sys intersects \ifextend four \fi lines of work: load balancing at the validator (ordering) level, adaptive partitioning and sharding, scheduling of ordered transactions across executors\ifextend{, and the use of time-derived signals under Byzantine faults}\fi.

\noindent \textbf{Validator level load balancing.} Stratus~\cite{gai2023scaling} targets the leader bottleneck of single-leader BFT via a shared mempool and proxy forwarding: overloaded replicas forward excess microblocks to a lightly loaded proxy. Neither mechanism carries over to certified DAG mempools~\cite{danezis2022narwhal}. Every validator is a full-time proposer whose blocks must reference batches processed by its own workers, leaving no idle capacity for proxy duty, and the $2f{+}1$ availability certificate must guarantee retrievability of batches, which Stratus's $f{+}1$ PAB shortcut cannot provide. Routing excess batches through a proxy therefore adds a relay step rather than removing load, and our implementation of this approach did not yield improvements.

Recent DAG-based protocols adapt to underperforming validators at the ordering-role level: Shoal, Shoal++, and HammerHead~\cite{spiegelman2024shoal,arun2025shoal++,tsimos2024hammerhead} use leader reputation to exclude slow or crashed leaders, while Sailfish, Mysticeti, Autobahn, and Mahi-Mahi~\cite{shrestha2024sailfish,babel2025mysticeti,giridharan2024autobahn,jovanovic2024mahi} reduce commit latency through multi-leader or uncertified designs. These works reassign consensus roles but leave the client-to-validator assignment, and hence the dissemination load, untouched. \sys is complementary: it rebalances client demand over any certified DAG mempool using metadata already in the DAG. Unlike systems that use time-derived signals to determine ordering~\cite{zhang2020byzantine,kelkar2020order,kelkar2023themis}, \sys's header timestamps and queue delays serve only as migration signals and never influence ordering.

\noindent \textbf{Adaptive partitioning and sharding.}
In trusted deployments, workload-driven partitioners~\cite{curino2010schism,quamar2013sword,taft2014estore,serafini2016clay} repartition or migrate data in response to workload skew. In untrusted settings, sharded BFT systems such as AHL~\cite{dang2019towards} and SharPer~\cite{amiri2021sharper} partition data across BFT clusters with a fixed mapping. BrokerChain~\cite{huang2022brokerchain} and Marlin~\cite{mehta2025adaptive} adaptively redistribute load. Marlin balances load across multiple BFT instances by changing data ownership, whereas \sys balances load within a single leaderless DAG-BFT instance by migrating client-to-validator assignments and committing migrations through the existing BFT log.
Marlin adapts despite Byzantine nodes that may report misleading statistics, using either hypergraph partitioning within a trusted administrative domain or decentralized key-affinity resharding.
The distinction is that \sys derives signals from the certified DAG rather than exchanged reports, and requires no separate atomic-commitment protocol for migrations.

\begin{table}[t]
\centering
\small
\setlength{\tabcolsep}{4pt}
\caption{Distributed transaction execution schemes along four
dimensions; $b$ is the batch size per consensus epoch, $m$ the number
of executor workers within a node, and $a$ the maximum
transaction size (number of states accessed).}
\label{tab:bg-compare}
\begin{tabular}{@{}lcccc@{}}
\toprule
 \textbf{System}         & \textbf{Bandwidth}        & \textbf{Lock time} & \textbf{Sched.} & \textbf{Load Bal.} \\
\midrule
LEAP~\cite{lin2016towards}          & $b$,   $1{\times}$ transfer  & Single-trip & per-tx$^{\dagger}$ & No       \\
PilotFish~\cite{kniep2025pilotfish} & $b/m$, $2{\times}$ transfer  & Round-trip  & $O(ba/m)^{\ddagger}$ & No     \\
Hermes~\cite{lin2021don}            & $b$,   $1{\times}$ transfer  & Single-trip & $O(b^{2}a^{2}m)$   & Yes      \\
\sys                                & $b$,   $1{\times}$ transfer  & Single-trip & $O(ba)$            & Yes      \\
\bottomrule
\end{tabular}
\vspace{2pt}
\\{\footnotesize $^{\dagger}$LEAP decides placement per transaction at
submission time rather than scheduling a batch.
$^{\ddagger}$Per worker sub-batch; placement is fixed by the object-to-worker partitioning.}
\end{table}

\noindent \textbf{Executor level load balancing.}
Table~\ref{tab:bg-compare} compares representative distributed execution mechanisms along bandwidth, lock-holding window, scheduling cost, and load balancing.
Of these, only Hermes~\cite{lin2021don} and \sys perform load-aware scheduling; \sys differs in preserving the committed order and supporting validators with different numbers of executor workers, motivating its full-batch decentralized scheduler with $O(ba)$ per-worker cost (\S\ref{sec:executor}). Deterministic databases fix execution order before locking: Calvin~\cite{thomson2012calvin} pioneered this, Bohm and early write visibility~\cite{faleiro2015rethinking,faleiro2017high} exploit the known order for multiversioning and pipelining, and Aria~\cite{lu2020aria} executes batches deterministically without prior read/write-set knowledge. Their performance still depends on partition quality, and none balances load across a BFT validator's executor workers.
\ifextend
LEAP~\cite{lin2016towards} pioneered what Hermes~\cite{lin2021don} later termed the look-present approach, migrating each distributed transaction's state to the executing node but deciding placement per transaction without any view of load. Hermes lifts it to the batch level with load-aware prescient routing, whose reordering of conflicting transactions a replicated BFT executor cannot adopt. PilotFish~\cite{kniep2025pilotfish}, the closest to our setting as a scale-out engine over an already ordered stream, fixes object-to-worker placement at dispatch time and leaves no handle for load-aware scheduling.

\noindent \textbf{Time-derived signals under Byzantine faults.}
Pompe~\cite{zhang2020byzantine} assigns each command the median of $2f{+}1$ signed local timestamps, which directly determines its position in the finalized order. Aequitas and Themis~\cite{kelkar2020order,kelkar2023themis} instead avoid timestamp aggregation and derive fairness constraints from replicas' relative receive orders, using Byzantine-robust agreement and threshold rules to constrain the admissible ordering. In all three systems, these validator observations enter the ordering semantics themselves. \sys draws the opposite boundary: header timestamps and queue delays never influence ordering. They serve only as performance signals for migration, filtered by medians and capped by certified throughput, so misreporting can degrade migration quality (\S\ref{sec:validator}) but can never affect ordering safety.
\fi
\section{Conclusion}
This paper addresses load imbalance across the ordering and execution layers of certified DAG-based BFT protocols. At the ordering layer, \sys dynamically rebalances client-validator assignments using load signals derived from the certified DAG, while a committed migration protocol preserves consistent assignments and serviceability under Byzantine faults. At the execution layer, \sys uses a deterministic, order-preserving scheduler to balance work across executor workers while minimizing data movement. Our evaluation on Narwhal and Tusk shows that \sys recovers performance under workload skew, validator heterogeneity, and shifting hotspots, scales across validators and executor workers, and incurs negligible overhead when load is balanced. End-to-end results further show that the two mechanisms compose effectively and that both layers must be addressed to remove their bottlenecks.

\balance

\bibliographystyle{ACM-Reference-Format}
\bibliography{_blockchain,_system,_privacy}

\end{document}